\documentclass[longauth]{aa} 

\usepackage{graphicx}
\usepackage{txfonts}
\usepackage{wasysym} 
\usepackage[colorlinks=true, allcolors=blue]{hyperref} 
\usepackage{threeparttable}
\usepackage{subcaption} 
\graphicspath{Figures/}
\begin{document}

   \title{A multiwavelength analysis of the spiral arms in the protoplanetary disk around WaOph 6}


   \author{S.~B.~Brown-Sevilla
          \inst{1}\fnmsep\thanks{Member of the International Max-Planck Research School for Astronomy and Cosmic Physics at the University of Heidelberg (IMPRS-HD), Germany}
          \and M. Keppler\inst{1}
          \and M. Barraza-Alfaro\inst{1}
          \and J. D. Melon Fuksman\inst{1}
          \and N. Kurtovic\inst{1}
          \and P. Pinilla\inst{1,2}
          \and M. Feldt\inst{1}
          \and W. Brandner\inst{1}
          \and C. Ginski\inst{3,4}
          \and Th. Henning\inst{1}
          \and H. Klahr\inst{1}
          \and R. Asensio-Torres\inst{1}
          \and F. Cantalloube\inst{1}
          \and A. Garufi\inst{5}
          \and R. G. van Holstein\inst{4}
          \and M. Langlois\inst{7,8}
          \and F. M\'enard\inst{9}
          \and E. Rickman\inst{10,11}
          \and M. Benisty\inst{12,9}
          \and G. Chauvin\inst{9,12}
          \and A. Zurlo\inst{13,14,8}
          \and P. Weber\inst{15,16,17}
          \and A. Pavlov\inst{1}
          \and J. Ramos\inst{1}
          \and S. Rochat\inst{9}
          \and R. Roelfsema\inst{18}
          }

   \institute{Max Planck Institute for Astronomy, K\"onigstuhl 17, 69117, Heidelberg, Germany\\
              \email{brown@mpia.de}
         \and
            Mullard Space Science Laboratory, University College London, Holmbury St Mary, Dorking, Surrey RH5 6NT, UK
         \and    
            Anton Pannekoek Institute for Astronomy, Science Park 904, NL-1098 XH Amsterdam, the Netherlands
         \and
         Leiden Observatory, Leiden University, P.O. Box 9513, 2300 RA Leiden, The Netherlands
         \and
         INAF, Osservatorio Astrofisico di Arcetri, Largo Enrico Fermi 5, 50125 Firenze, Italy
         \and
         European Southern Observatory, Alonso de Córdova 3107, Casilla 19001, Vitacura, Santiago, Chile
         \and
         CRAL, UMR 5574, CNRS, Universit\'e de Lyon, Ecole Normale Sup\'erieure de Lyon, 46 all\'ee d’Italie, 69364 Lyon Cedex 07, France
         \and
         Aix Marseille Univ., CNRS, CNES, LAM, Marseille, France
         \and
         Univ. Grenoble Alpes, CNRS, IPAG, 38000 Grenoble, France
         \and
         Geneva Observatory, University of Geneva, Chemin des Maillettes 51, CH-1290 Sauverny, Switzerland
         \and
         European Space Agency (ESA), ESA Office, Space Telescope Science Institute, 3700 San Martin Drive, Baltimore, MD 21218, USA
         \and
         Unidad Mixta Internacional Franco-Chilena de Astronom\'ia (CNRS, UMI 3386), Departamento de Astronom\'ia, Universidad de Chile, Camino El Observatorio 1515, Las Condes, Santiago, Chile
         \and
         N\'ucleo de Astronom\'ia, Facultad de Ingenier\'ia y Ciencias, Universidad Diego Portales, Av. Ejercito 441, Santiago, Chile
         \and
         Escuela de Ingenier\'ia Industrial, Facultad de Ingenier\'ia y Ciencias, Universidad Diego Portales, Av. Ejercito 441, Santiago, Chile
         \and
         Departamento de Astronomía, Universidad de Chile, Camino El Observatorio 1515, Las Condes, Santiago, Chile
         \and
         Universidad de Santiago de Chile, Av. Libertador Bernardo O’Higgins 3363, Estaci\'on Central, Santiago, Chile
         \and
         Center for Interdisciplinary Research in Astrophysics and Space Exploration (CIRAS), Universidad de Santiago de Chile, Estaci\'on Central, Chile
         \and
         NOVA Optical Infrared Instrumentation Group, Oude Hoogeveensedijk 4, 7991 PD Dwingeloo, The Netherlands
             }

   \date{Received March 11, 2021 ; accepted July 13, 2021 }   

 
  \abstract
   {In recent years, protoplanetary disks with spiral structures have been detected in scattered light, millimeter continuum, and CO gas emission. The mechanisms causing these structures are still under debate. A popular scenario to drive the spiral arms is the one of a planet perturbing the material in the disk. However, if the disk is massive, gravitational instability is usually the favored explanation. Multiwavelength studies could be helpful to distinguish between the two scenarios. So far, only a handful of disks with spiral arms have been observed in both scattered light and millimeter continuum.} 
   {We aim to perform an in-depth characterization of the protoplanetary disk morphology around WaOph 6 analyzing data obtained at different wavelengths, as well as to investigate the origin of the spiral features in the disk.}
   {We present the first near-infrared polarimetric observations of WaOph 6 obtained with SPHERE at the VLT and compare them to archival millimeter continuum ALMA observations. We traced the spiral features in both data sets and estimated the respective pitch angles. We discuss the different scenarios that can give rise to the spiral arms in WaOph 6. We tested the planetary perturber hypothesis by performing hydrodynamical and radiative transfer simulations to compare them with scattered light and millimeter continuum observations.}
   {We confirm that the spiral structure is present in our polarized scattered light $H$-band observations of WaOph 6, making it the youngest disk with spiral arms detected at these wavelengths. From the comparison to the millimeter ALMA-DSHARP observations, we confirm that the disk is flared. We explore the possibility of a massive planetary perturber driving the spiral arms by running hydrodynamical and radiative transfer simulations, and we find that a planet of minimum 10 M$_{\rm{Jup}}$ outside of the observed spiral structure is able to drive spiral arms that resemble the ones in the observations. We derive detection limits from our SPHERE observations and get estimates of the planet's contrast from different evolutionary models.}
   {Up to now, no spiral arms had been observed in scattered light in disks around K and/or M stars with ages $<1$\,Myr. Future observations of WaOph 6 could allow us to test theoretical predictions for planet evolutionary models, as well as give us more insight into the mechanisms driving the spiral arms.}


   \maketitle
%

\section{Introduction}
\label{sec:intro}

The study of protoplanetary disks provides insight into our understanding of the formation and early evolution of planets. Modern observational techniques have enabled significant progress to be made in this task. Recent observations in both scattered light and millimeter continuum have shown the striking frequency with which these disks present structures, such as gaps, rings, or spirals \citep[e.g.,][]{Avenhaus2018, Long2018, Andrews2020, Cieza2020}. 

In particular, spiral arms have been observed in more than a dozen young disks, spanning from tens to hundreds of au. They have been found in scattered light \citep[e.g.,][]{Muto2012, Wagner2015, Benisty2015, Stolker2016, Garufi2020, Muro-Arena2020}, millimeter continuum \citep[e.g.,][]{Perez2016, Huang2018III, Rosotti2020}, and more recently in CO gas emission \citep[e.g.,][]{Tang2017, Kurtovic2018}. However, for the disks that have been imaged at different wavelengths, some of these spirals are only present in either scattered light, and not in the millimeter continuum, or vice versa. The mechanisms that drive spiral arms in these disks are probably manifold, and they are a matter of debate in many individual cases. 

Considering that planets are born in protoplanetary disks, the observed structures have been frequently linked with the presence of planets forming within the disk \citep[e.g.,][]{Muto2012, Pohl2015, Dong2018, Calcino2020, Ren2020}. These planets perturb the disks via gravitational interactions, and these perturbations can cause the formation of spirals. In these cases, the planets cause spiral arms both in the interior and exterior of their own orbit. In particular, when the planet is fairly massive, it can trigger a secondary spiral arm on the inside of its orbit. For a planet of tens of M$_{\mathrm{Jup}}$, the primary and secondary spiral arm reach approximately a $m=2$\footnote{$m$ is the azimuthal mode number, which represents the number of spiral arms.} \citep{Dong2015}. From such planet-driven spirals, we can study the mass and location of the potential planets. Another mechanism that can drive spiral arms is gravitational instability \citep[GI, e.g.,][]{Goldreich1965}. This happens when the self-gravity perturbations in the disk dominate over the restoring forces of gas pressure and differential rotation \citep{Toomre1964}, which can be translated to the disk being gas rich and relatively massive with respect to its host star (M$_{\mathrm{disk}}/{\mathrm{M}_{*}} \gtrsim 0.1$ \cite{Kratter2016}).
Spiral arms generated by GI can allow one to better constrain the disk mass. A third possible cause for the presence of spiral structure is a flyby event of a close companion to the system \citep[e.g.,][]{Cuello2019, Menard2020} that perturbs the disk material causing over densities that form the spiral pattern. And finally, a combination of some of the physical processes described above has also been investigated \citep[e.g.,][]{Pohl2015}.

Distinguishing between the possible physical processes driving spiral structure in protoplanetary disks is not an easy task. In fact, measuring the disk gas mass alone is already challenging. Having a more comprehensive understanding of these disks can be useful to decipher between the physical processes that are taking place within them. One approach is to use multiwavelength observations to trace their different components. In particular, images obtained in scattered light are sensitive to micron-sized dust grains at the disk surface, which under typical disk conditions are very well coupled to the gas. Images at (sub)millimeter wavelengths, on the other hand, trace larger grains within the disk midplane, and they are less well coupled to the gas. Therefore, comparing images at these wavelengths with similar angular resolution can potentially reveal the different morphologies of different disk components. Actually, it may be possible to distinguish between planet- or GI-induced spirals by comparing scattered light and millimeter continuum observations, since dust trapping in spiral arms is likely to be more efficient in gravitationally unstable disks \citep{Juhasz2015, Dipierro2015}. Previous studies have utilized observations in both near-infrared (NIR) and millimeter continuum to analyze the disks' morphology, as well as the gas and dust distributions within them \citep[e.g.,][]{vanBoekel2017, Rosotti2020}, however, so far there are only a handful of such works.

\begin{figure}
   \centering
   \includegraphics[width=0.4\textwidth]{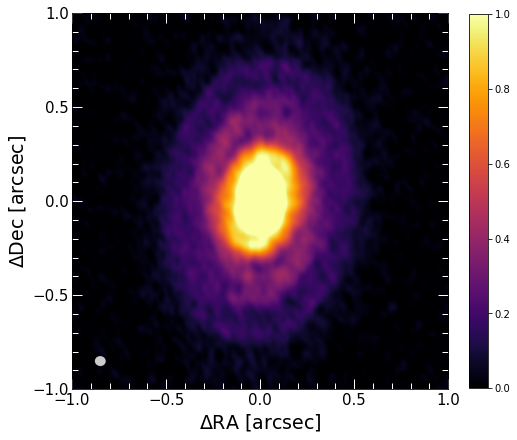}
   \caption{ALMA 1.25 mm continuum image of WaOph 6 from the ALMA-DSHARP survey \protect\citep{Huang2018III}. The beam size is shown in the lower left corner.} 
    \label{fig:alma}
    \end{figure}

In this paper, we present NIR polarimetric observations of WaOph 6 obtained using the VLT/SPHERE instrument \citep{Beuzit2019}. These observations have a spatial resolution of $\sim$51\,mas ($\sim$ 6\,au), similar to that of the recent ALMA-DSHARP millimeter wavelength observations \cite[][see Fig.~\ref{fig:alma}]{Andrews2018}. We report a NIR scattered light counterpart of the innermost spiral arms reported in \cite{Huang2018III}. Additionally, we explore one of the scenarios that can give rise to the spiral pattern observed in the disk.

The paper is structured as follows: in Section~\ref{sec:waoph6} we introduce WaOph 6 and the previous studies on its disk. In Section~\ref{sec:obs} we describe our scattered light observations and data reduction procedure. Our results are presented in Section~\ref{sec:results}. We describe the modeling setup and the comparison between the simulations and the observations on Section~\ref{sec:model}. In Section~\ref{sec:disc} we discuss our results, and finally in Section~\ref{sec:sum} we summarize our findings and list our conclusions.
   
\section{Stellar and disk properties}
\label{sec:waoph6}

WaOph6 is a K6 star \citep{Eisner2005}, and member of the Ophiuchus moving group at a distance of 122.5 $^{+0.3}_{-0.2}$ pc \citep{Gaia2020} located near the L162 dark cloud. It was first identified as a suspected T Tauri star by \cite{Henize1976}, and then confirmed by \cite{Walter1986}. Here we constrain the stellar mass and age based on the updated photometry and Gaia parallax. We retrieved the full spectral energy distribution (SED) from Vizier\footnote{\url{http://vizier.unistra.fr/vizier/sed/}} and employed a Phoenix model of the stellar photosphere \citep{Hauschildt1999} with effective temperature T$_\mathrm{eff}=4200$~K \citep{Eisner2005}, surface gravity log(g)=-4.0, and an optical extinction $A_V = 2.8 \pm 0.3$ mag calculated from the V, R, and I photometric fluxes. We integrated the stellar model scaled to the average V magnitude and Gaia distance of 122.5 pc obtaining a stellar luminosity of L$_*=1.91 ^{+0.70}_{-0.51}$ L$_{\odot}$. Then, we placed the source on the HR diagram and constrain a stellar mass M$_*=0.7 \pm 0.1$ M$_{\odot}$ and an age $t=0.6 \pm 0.3$ Myr through different sets of PMS tracks \citep[Parsec, MIST, Baraffe;][]{Bressan2012, Choi2016, Baraffe2015} with error bars propagated from L$_*$ and T$_\mathrm{eff}$ ($\pm$ 100 K).

The disk around WaOph 6 has been a common target for millimeter continuum surveys looking to constrain and characterize the structure in protoplanetary disks \citep[e.g.,][]{AndrewsWilliams2007, Andrews2009, Ricci2010, Andrews2018}. Submillimeter Array (SMA) observations were used along with a parametric model to constrain density structure parameters \citep{Andrews2009}. The disk model that best fitted the thermal continuum data and spectral energy distribution (SED) was that of a flat cold disk with a total disk mass (gas $+$ dust-to-gas ratio of 1:100) of 0.077\,$M_\odot$. With observations from the Australia Telescope Compact Array (ATCA), \cite{Ricci2010} analyzed and modeled the SED of WaOph~6, adopting a distance of $\sim$130\,pc and an outer radius (R$_{out}$) interval of $175-375$\,au, and they find dust mass estimates (M$_\mathrm{dust}$) between $8\times10^{-5}\,\mathrm{M}_{\odot}$ and $9.8\times10^{-5}\,\mathrm{M}_\odot$, depending on the assumed dust size distribution power-law index ($q=2.5$ or $q=3$). More recently, WaOph~6 was observed by ALMA within the DSHARP program \citep[Disk Substructures at High Angular Resolution Project,][]{Andrews2018}. These millimeter continuum observations showed that the disk has a set of symmetric spiral arms that extend to $\sim$70\,au, a gap at 79\,au and a bright ring at 88\,au \citep{Huang2018III}. The disk has an inclination ($i$) of 47.3$^{\circ}$ and a position angle (PA) of 174.2$^{\circ}$ obtained from ellipse fitting on the dust continuum emission \citep[see][for more details]{Huang2018}, and gas observations have shown that it suffers from mild molecular cloud contamination \citep{Reboussin2015}. We summarize the stellar and disk physical parameters in Table~\ref{tab:wo6}, where we include the different values for the disk mass found in the literature, as well as our own M$_\mathrm{dust}$ estimate obtained following the procedure described in Appendix \ref{app:a}. 

Up to now, only seven disks have been known to have spiral arms in millimeter continuum wavelengths: WaOph~6, Elias 27, IM Lup, HT Lup A, AS 205 N, MWC 758, and HD100453 \citep{Huang2018III, Kurtovic2018, Dong2018, Rosotti2020}, and only the first three are single systems. Out of these three, only IM Lup has published polarized scattered light observations \citep{Avenhaus2018}, however, with no spiral arms visible at these wavelengths. 

\noindent
\begin{table}
\caption[WaOph 6 parameters from the most recent literature]{Stellar and disk parameters of WaOph 6 from the most recent literature.} 
\begin{threeparttable}
	\centering
	\begin{tabular}{ccccccccccccccccccccc}
	\hline
    \hline
    	\textbf{Stellar parameters} & Value & Ref. \\
        \hline
        Spectral type  & K6 & a \\
		Age  & 0.7 Myr & a \\
		Distance \textit{d} & 122.5 $\pm$ 5 pc & b \\
		Mass & 0.9 M$_\odot$ & a \\
		Radius & 2.8 R$_\odot$ & a \\
        Temperature & 4205 $K$ & a \\
        Visual magnitude ($V$ band) & 13.3 $\pm$ 0.01 mag & c \\
        \hline
        \textbf{Disk properties} \\
        \hline
        Inclination \textit{i}  & 47.3$^{\circ}$ & d \\
		Position angle (PA)  & 174.2$^{\circ}$ & d \\
	    Gas mass\footnotemark[1] & 7.7x$10^{-2}$ M$_\odot$ & e \\
	    Dust mass\footnotemark[2] & 8x$10^{-5}$ M$_\odot$ & a \\
	    Dust mass & 1.4x$10^{-4}$ M$_\odot$ & f \\
        \hline
	\end{tabular}  
    \begin{tablenotes}
	    \item [1] {\footnotesize Assuming a gas-to-dust ratio of 100:1 and based on SMA SED modeling.}
	    \item [2] {\footnotesize Obtained with a power-law index for the grain size distribution $q = 2.5$.}
	    \item {\footnotesize \textbf{Notes.} \hspace{1em} a) \protect\cite{Ricci2010}, b) \protect\cite{Gaia2020}, c) \protect\cite{Zacharias2012}, d) \protect\cite{Huang2018}, e) \protect\cite{Andrews2009}, f) This work} 
	\end{tablenotes}
\end{threeparttable}
\label{tab:wo6}
\end{table}

\begin{table}
\caption[Observing log]{Log of observations.} 
	\centering
	\begin{tabular}{lc}
	\hline
    \hline
        Date  & 21-06-2018 \\
		Filter  & $H$-band (1.625 $\mu$m) \\
	    UT start/end & 01:58:36/02:24:30  \\
	    Exposure time & 96 s \\
	    Airmass & $\sim$1.0 \\
	    Seeing & $\sim$ 0.5" \\
	    Coherence time ($\tau_0$) & $\sim$ 4 ms \\
	    Wind speed & $\sim$ 3.8 m/s \\
	    Total exposure time & $\sim$ 1500 s \\
        \hline
 	\end{tabular}  
\label{tab:obslog}
\end{table}

\section{Observations and data reduction}
\label{sec:obs}

WaOph 6 was observed with the VLT/SPHERE high-contrast instrument \citep{Beuzit2019} within the DISK/SHINE \citep[SpHere INfrared survey for Exoplanets,][]{Chauvin2017} Guaranteed Time Observations (GTO) program on the night of June 21, 2018 (see Table \ref{tab:obslog}). The observations were carried out with the IRDIS Dual-beam Polarimetric Imaging (DPI) mode \citep{Langlois2014, deBoer2020, vanHolstein2020} in $H$-band \citep[$\lambda_c$=1.625\,$\mu$m; $\Delta\lambda$=0.291\,$\mu$m; where $\lambda_c$ denotes the central wavelength and $\Delta\lambda$ denotes the full width at half maximum (FWHM) of the filter transmission curve; pixel scale 12.25\,mas/px,][]{Maire2016} in field stabilized mode using an apodized Lyot coronagraph, having a focal plane mask of 93\,mas radius \citep{Carbillet2011}. A total of four polarimetric cycles were recorded, with 96\,s of integration time per exposure, resulting in a total integration time of about 25 minutes. Each polarimetric cycle consisted of adjusting the half-wave plate (HWP) at four different switch angles: 0$^{\circ}$, 45$^{\circ}$, 22.5$^{\circ}$, and 67.5$^{\circ}$. At each HWP position the two orthogonal linear polarization states are measured simultaneously, resulting in eight images per cycle, corresponding to the Stokes components: $(I\pm Q)/2$, $(I\mp Q)/2$, $(I\pm U)/2$, and $(I\mp U)/2$. To obtain the Stokes components $Q^+$, $Q^-$, $U^+$ and $U^-$, one orthogonal state is subtracted from the other at each of the HWP angles. Besides the science data, star center frames at the beginning and end of the sequence, as well as flux calibration frames were obtained. For the star center frames, the deformable mirror (DM) waffle mode was used \citep[see][for more details on this mode]{Langlois2013}. Two flux calibration frames (images of the target star without the coronagraph) were obtained with an exposure time of 2\,s and a neutral density (ND1) filter to prevent saturation. 
We measure a point spread function (PSF) FWHM of $\sim$51\,mas by fitting a Gaussian function to the flux frames. The weather conditions were stable during the observations with a seeing of $\sim$0.5", a coherence time ($\tau_0$) of $\sim$4\,ms, and wind speed of $\sim$3.8\,m/s. The Strehl ratio was about 0.7, however, the low scattered light intensity resulted in a rather low signal-to-noise ratio (S/N).

For the data reduction, we used the IRDAP pipeline\footnote{\href{https://irdap.readthedocs.io}{https://irdap.readthedocs.io}} version 1.3.2. \citep{vanHolstein2020}. First, the pipeline preprocesses the data by performing the usual sky background subtraction, flat fielding, bad-pixel identification and interpolation, and star centering corrections. Subsequently, polarimetric differential imaging (PDI) is performed by applying the double-sum and double-difference method described in \cite{deBoer2020} to obtain a set of Stokes $Q$ and $U$ frames. Finally, the data are corrected for instrumental polarization and crosstalk effects by applying a detailed Mueller matrix model of the instrument \cite[see][for more details on the data reduction procedure]{vanHolstein2020}, yielding the final $Q$ and $U$ images. The final PDI images are corrected for true north following the procedure established by \cite{Maire2016}. IRDAP then obtains the linearly polarized intensity ($PI$) image using the final $Q$ and $U$ images, from

\begin{equation}
      PI = \sqrt[]{Q^2 + U^2}.
\end{equation}

\noindent Next, the pipeline computes the azimuthal Stokes parameters following \citep{deBoer2020}:  

\begin{equation}\label{eq:QphiUphi}
\begin{aligned}
      Q_\phi = -Q\cos{(2\phi)} - U\sin{(2\phi)},\\
      U_\phi = +Q\sin{(2\phi)} - U\cos{(2\phi)},
\end{aligned}      
\end{equation}

\noindent where $\phi$ is the position angle (PA) measured east of north with respect to the position of the star. In the definition above, a positive signal in the $Q_\phi$ image corresponds to a signal that is linearly polarized in the azimuthal direction, while a negative signal denotes radially polarized light in $Q_\phi$. $U_\phi$ contains any signal polarized at $\pm45^{\circ}$ with respect to the radial direction. This means that for disks with low inclinations (i.e., $i < 45^{\circ}$), almost all of the scattered light is expected to be included as a positive signal in $Q_\phi$, while the $U_\phi$ image can be considered as an upper limit of the noise level. In the case of WaOph~6, we expect some physical signal in the $U_\phi$ image due to the inclination of the disk (see Table \ref{tab:wo6}). The resulting $Q_\phi$ and $U_\phi$ images are shown in Appendix~\ref{app:sph_img}. 


\section{Results} 
\label{sec:results}

\begin{figure*}
    \centering
    \includegraphics[width=0.95\textwidth]{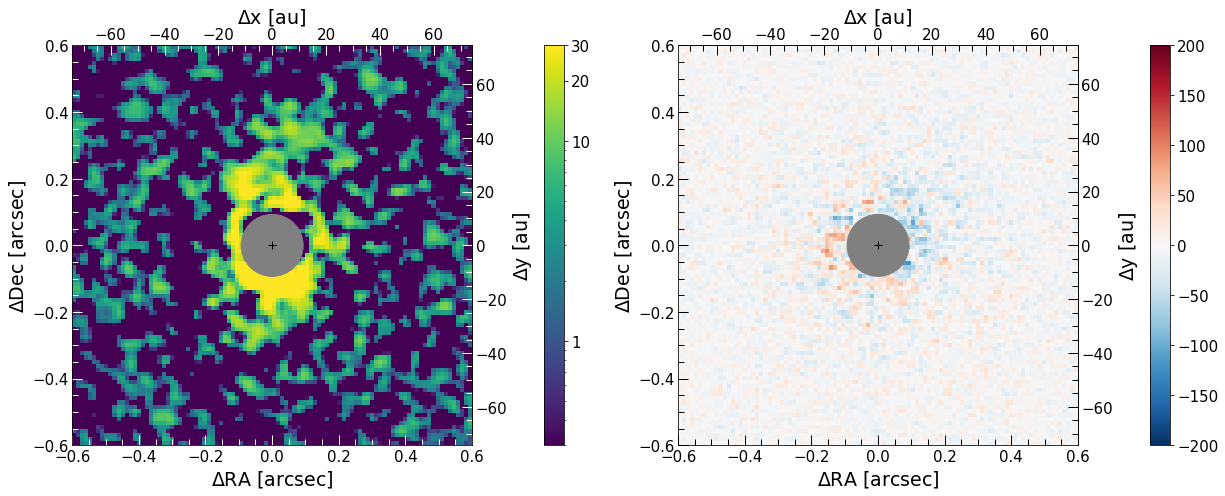}
      \caption{\textit{Left:} Close up of the final $Q_{\phi}$ SPHERE/IRDIS-DPI image after removing low frequency structures (see text for details) and applying a Gaussian kernel of size 0.1 $\times$ FWHM to smooth the images and enhance the spiral features. \textit{Right:} Close up of the final $U_{\phi}$ image showing the positive and negative signal. The 93\,mas coronagraph is indicated by the gray circle, and the cross indicates the position of the star. The flux is normalized to the maximum value in the $Q_{\phi}$ image. 
              }
         \label{fig:finalqphi}
   \end{figure*}
   
   \begin{figure}
    \centering
    \includegraphics[width=0.4\textwidth]{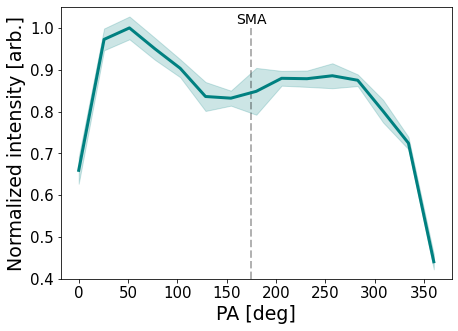}
      \caption{Azimuthal profile of the deprojected Q$_{\phi}$ image, radially averaged over a ring of 0.22-0.29" ($\sim$27 - 36\,au). The two peaks at around 50$^\circ$ and 210$^\circ$ correspond to the launch locations of the two spiral arms. The error is taken to be 2 $\sigma$. The dashed line indicates the location of the disk's semi-major-axis. 
              }
         \label{fig:azimuthalplot}
   \end{figure}

    \subsection{SPHERE/IRDIS-DPI observations}
    \label{sec:sph_obs}

    In Fig. \ref{fig:finalqphi} we show the final, processed $Q_{\phi}$ and $U_{\phi}$ images. Due to the low S/N, we sharpened the images by subtracting a version of them which was convolved by a Gaussian kernel with the size of 10 pixels, which removes low frequency structures, and then we convolved them with a Gaussian kernel of size 0.1 $\times$ FWHM to smooth the images in order to enhance the spiral features. We observe the launch of the spiral arms up to $\sim$0.3" (40\,au), as seen in the $Q_{\phi}$ image (Fig. \ref{fig:finalqphi}, left). As mentioned in Section \ref{sec:obs}, the $U_{\phi}$ image (Fig. \ref{fig:finalqphi}, right) in this case contains almost no signal and can be used as an upper limit of the noise level. For a better visualization of the spiral features, we plotted the azimuthal profile by first deprojecting the filtered $Q_{\phi}$ image, and taking the average flux within the ring between $\sim$27 and 36\,au in azimuthal bins of 15$^\circ$. The distance range is chosen due to the presence of the coronagraph at lower radii, and the S/N decrease at higher radii. In Fig. \ref{fig:azimuthalplot}, we show the smoothed azimuthal profile. The spiral arms are seen as the two peaks between $20^\circ - 100^\circ$ and $200^\circ - 310^\circ$. To estimate the spiral arm intensity contrast between the spiral and inter-spiral regions, we measured the peak intensities of the spiral arms in radial bins spaced by 2\,au. 
    We estimated the inter-spiral region intensity by taking the minimum value in each bin. The contrast is then the ratio between the peak intensities and the inter-spiral intensities. On average, we find the spiral arm contrast to be 1.5. 
    
    \subsection{Companion candidate}
    \label{sec:cc}
   
   \begin{figure}
    \centering
    \begin{subfigure}[]{1\textwidth}
        \includegraphics[width=0.455\linewidth]{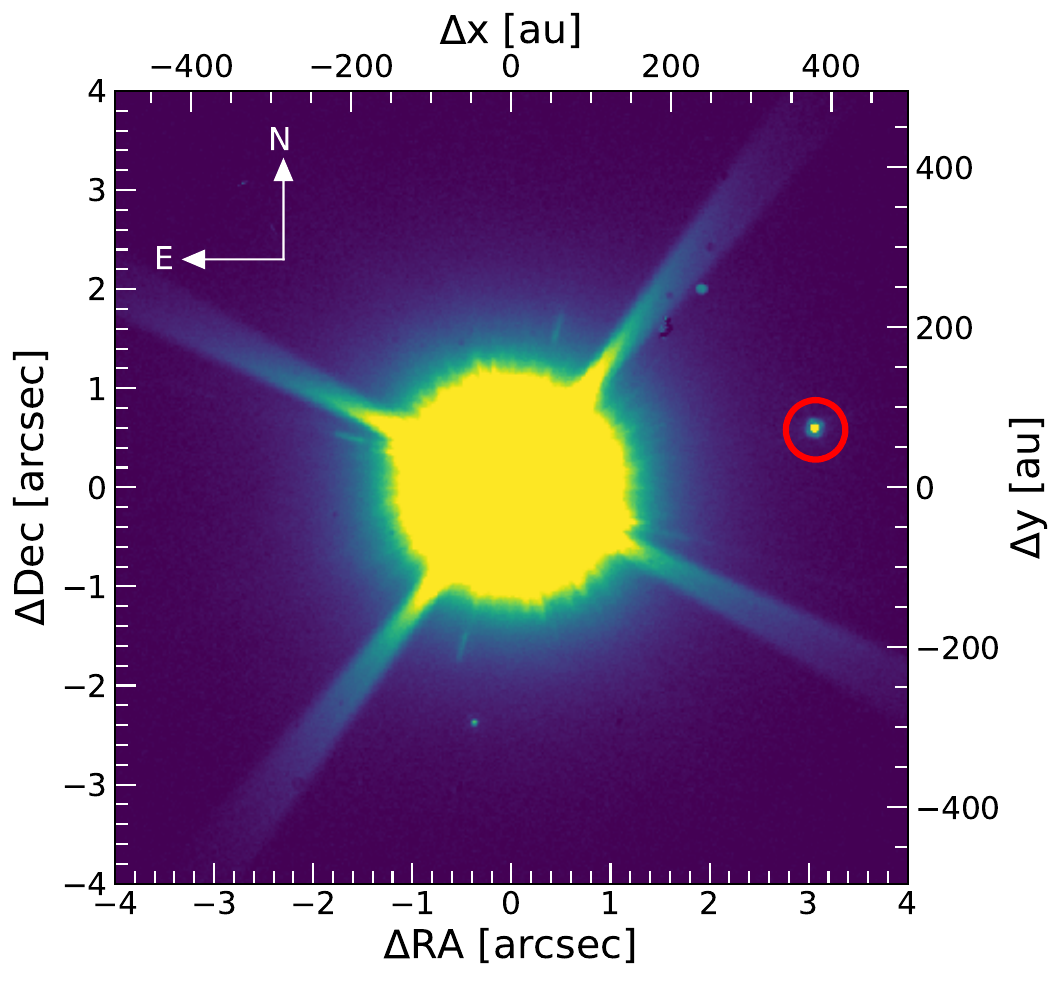}\\ 
    \end{subfigure}
    
    \begin{subfigure}[]{1\textwidth}
        \includegraphics[width=0.4\linewidth]{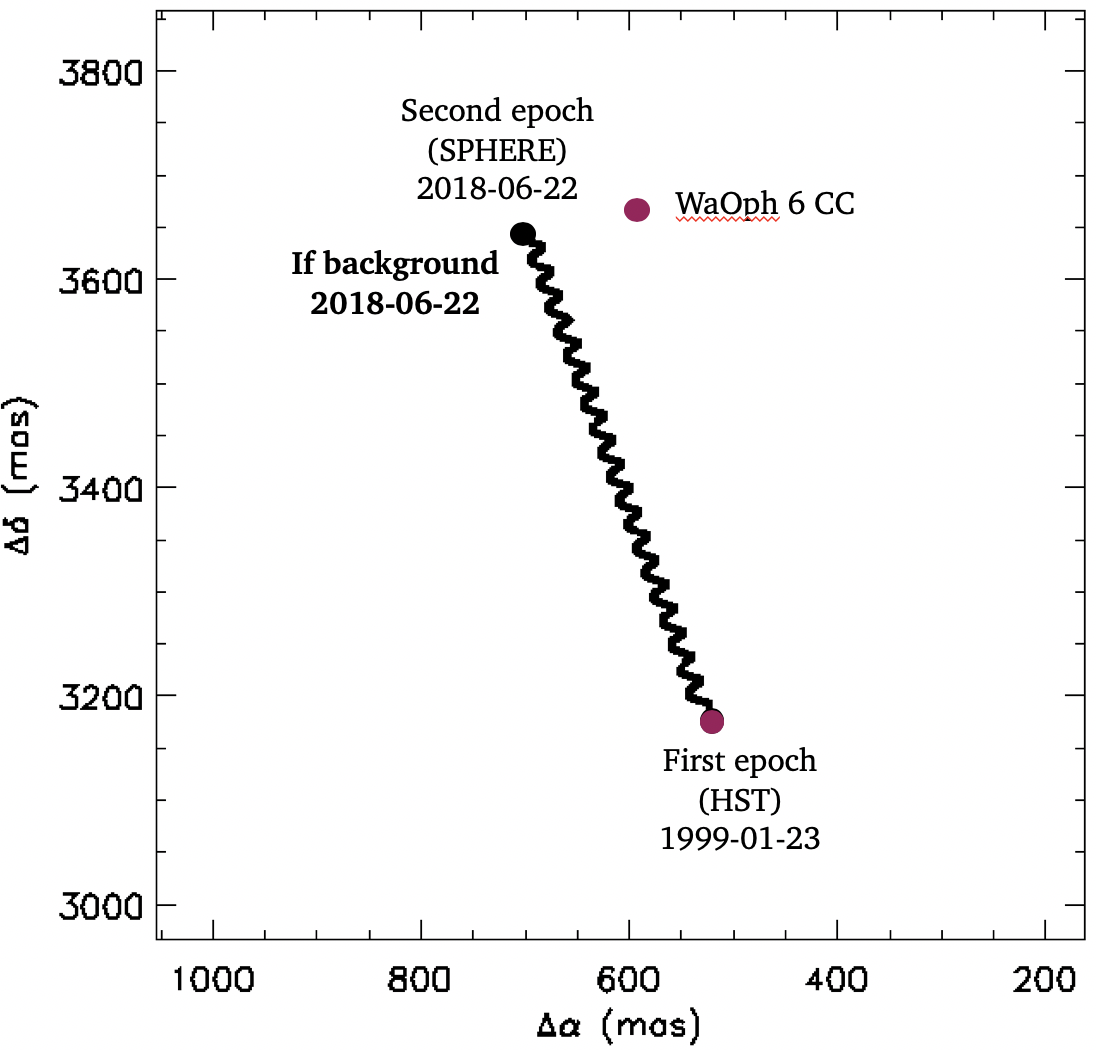}
    \end{subfigure}
    \caption[Astrometry analysis]{\textit{Top:} Total intensity SPHERE/IRDIS-DPI image of WaOph 6. Encircled in red is the CC. \textit{Bottom:} Astrometry plot of the CC of WaOph 6. The markers show the position of the CC, at the initial HST epoch, and further in time at the SPHERE epoch. The black curve traces the path a stationary background object would have followed relative to WaOph~6 between the two epochs. 
    }
    \label{fig:cc}
   \end{figure}

   We detect a companion candidate (CC) in our data as shown in the total intensity image in the top panel of Fig. \ref{fig:cc}, where the CC is more visible than in the polarized light frames. The CC is located at a projected distance of $\sim$400\,au ($\sim$3") and has a brightness contrast of 10$^{-3}$ with respect to WaOph 6. We used archival HST data (from 1999-01-23) as an additional epoch to perform an astrometric analysis in order to verify if the CC is bound to the system. The resulting astrometry plot is shown in the bottom panel of Fig.~\ref{fig:cc}, where the black curve traces the path a stationary background object would have followed relative to WaOph 6 between the two epochs, and the markers show the position of the CC at both the HST and the SPHERE epochs. Since the CC is located near the final position a background object would be located at, we conclude that the object is not bound to the system, and therefore could not be considered as an external perturber causing the spirals. We note that the CC is not reported in the Gaia EDR3, despite of its presence in the two data sets described above.

   \subsection{Comparison to ALMA Observations}
   \label{sec:alma}
   
   \begin{figure}
   \centering
   \includegraphics[width=0.46\textwidth]{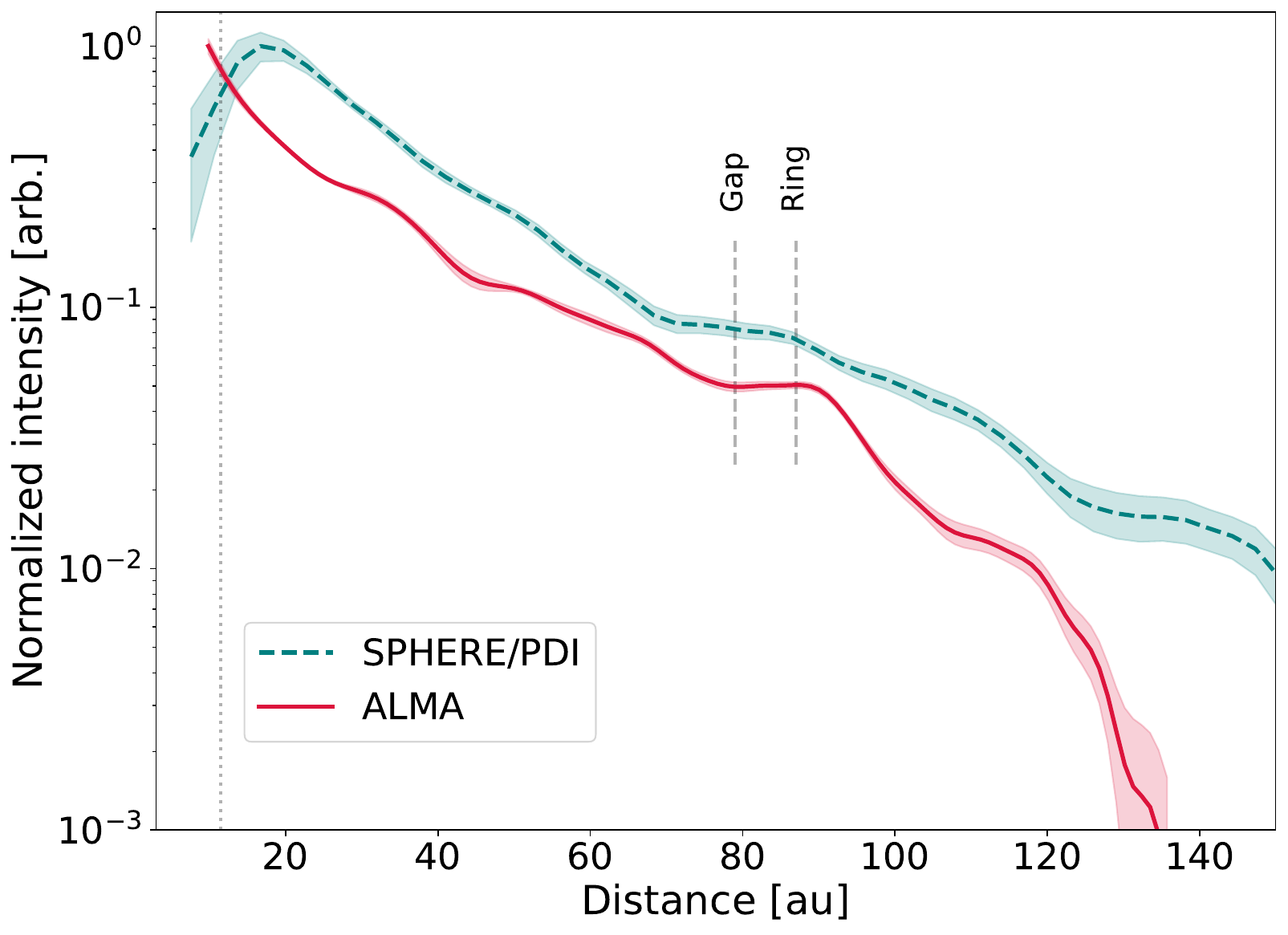} 
      \caption{Radial intensity profiles of both the SPHERE and ALMA images of WaOph 6. Plotted on a logarithmic scale and normalized to the maximum intensity of each image. The SPHERE profile is taken from the reduced $Q_{\phi}$ image with an applied Gaussian kernel of size 0.1 $\times$ FWHM to smooth the curve, and it is shown up to 3$\sigma$ of the intensity. The dotted line indicates the coronagraph coverage.}
         \label{fig:radp}
   \end{figure}

    WaOph 6 was observed within ALMA/DSHARP \citep{Andrews2018} in Band~6, at a frequency of 239 GHz (1.3 mm). Observations at these wavelengths sample the millimeter-sized dust grains that are typically located in the disk midplane \citep[see e.g.,][]{Villenave2020}. On the other hand, our SPHERE observations trace the light scattered from submicron sized dust grains located at the disk surface which are typically well-coupled to the gas. In this section we do a first comparison of the two data sets.
    
    \subsubsection{Radial profiles}

    In Fig.~\ref{fig:radp} we plot the radial intensity profiles of the SPHERE/IRDIS-DPI H-band image (teal curve), and the ALMA image in Fig.~\ref{fig:alma} (crimson curve). The curves have been normalized to the maximum intensity value on each image for visualization purposes. In order to obtain a smooth profile, we apply a Gaussian kernel of size 0.1 $\times$ FWHM to the reduced $Q_{\phi}$ image in Appendix~\ref{app:sph_img}. We obtained the profiles by taking the azimuthal average of the image intensity in rings of radius 3\,au. For this we considered the latest literature values for $i$ and PA (listed on Table \ref{tab:wo6}), and we use the \texttt{aperture\_photometry()} function from the \texttt{photutils} python package, which allows to perform aperture photometry within elliptical annuli. This permits to get the radial profile without first deprojecting the image. 
    
    As expected from the fact that the two images trace different dust sizes, the profiles do not perfectly overlap. A closer look between 70\,au and 90\,au shows that the substructures created by the gap and the ring described in Section \ref{sec:waoph6}, are present in both profiles, with a slight shift. 
    The initial drop in the intensity of the SPHERE/IRDIS-DPI profile is due to the use of a coronagraph in these observations. The cut in the ALMA data profile can be attributed to the emission from the large grains being limited to the central $\sim$130 au of the disk.
    
\begin{table}
\caption[Spiral fit parameters]{Spiral pitch angles for the protoplanetary disk around WaOph 6.} 
	\centering
	\begin{tabular}{ccccccccccccccccccccc}
	\hline
    \hline
    	Source & Spiral arm & Log. $\mu$ [$^{\circ}$] & Arch. $\mu$ [$^{\circ}$] \\ 
        \hline \vspace{0.5em}
        SPHERE  & N & 19.79$_{-0.11}^{+0.12}$ & 19.54 \\ \vspace{0.5em} 
          & S & 14.04$_{-0.06}^{+0.07}$ & 16.05 \\ \vspace{0.5em}
		ALMA  & N1 & 13.49$_{-0.19}^{+0.30}$ & 18.97 \\ \vspace{0.5em} 
		  & S1 & 18.26$_{-0.07}^{+0.04}$ & 15.98 \\ \vspace{0.5em}
		  & N2 & 10.75$_{-0.15}^{+0.23}$ & 17.86 \\ \vspace{0.5em}
		  & S2 & 9.09$_{-0.10}^{+0.21}$ & 15.26 \\ 
        \hline
	\end{tabular} 
	\begin{tablenotes}
	    \item {\footnotesize \textbf{Note:} For the SPHERE data, N corresponds to the northern spiral and S to the southern spiral. In the case of the ALMA data, N1 corresponds to the northern inner spiral, S1 to the southern inner spiral, N2 to the northern outer spiral and S2 to the southern outer spiral. For the Archimedean fit, the pitch angles are estimated at 35\,au.} 
	\end{tablenotes}
\label{tab:pitch}
\end{table}
   
   \begin{figure*}
    \centering
    \includegraphics[width=0.9\textwidth]{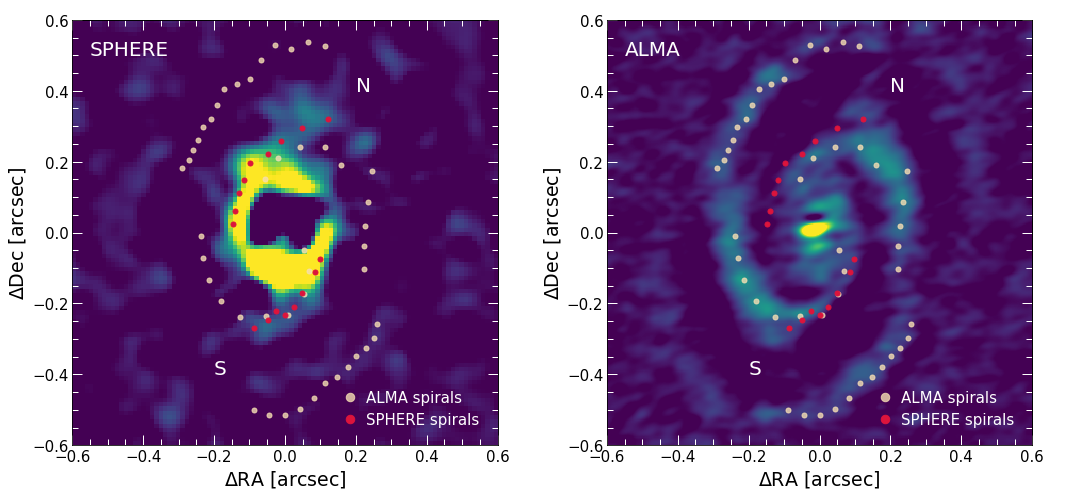}
      \caption{\textit{Left:} SPHERE $Q_{\phi}$ image as described in Fig. \ref{fig:finalqphi} but with a Gaussian kernel of size 0.2 $\times$ FWHM with the spiral arms retrieved from both the ALMA and this image overplotted. \textit{Right:} ALMA continuum image generated as described in the Appendix~\ref{app:alma}, with the overplotted spiral arms retrieved from both the SPHERE $Q_{\phi}$ and this image. The ``N'' and ``S'' indicate the northern and southern spirals, respectively.}
    \label{fig:sphspirals}
   \end{figure*}
    
    \subsubsection{Spiral arms}
    
    Next, we performed a spiral search on each data set to compare the location of the spiral arms at both wavelengths. In the case of the SPHERE data, the spiral features are better seen when the image is plotted in polar coordinates, where spiral arms appear as inclined lines. To obtain the polar plot we first deprojected the image and then converted to polar coordinates. We then used the python function \texttt{peak\_local\_max} from the \texttt{skimage} package to search for the peak emission points around the location of the spiral arms. 
    To trace the spirals in the ALMA data we used two different images generated using the \texttt{tclean} task in CASA 5.4.1 \citep{McMullin2007}. We used uv-taper = [`0.010arcsec', `0.010arcsec', `10deg'], and two different robust values (-1.0 and 1.0) to generate an image with higher resolution in the central and in the outer regions, respectively. 
    Then, we searched for peak emission points along the radial direction and took the 3$\sigma$ emission ones as the spiral arms. We repeated this process in both of the images described above. The resulting spirals are shown in Fig. \ref{fig:sphspirals}, where we overplot the retrieved spiral arms on both the SPHERE (left) and the ALMA image (right). To obtain the SPHERE image we followed the same procedure described in Section \ref{sec:results} with a Gaussian kernel of size 0.2 $\times$ FWHM. We did this to match the scaling of the ALMA image for better comparison purposes. To generate the ALMA image, we used the \texttt{frank} \citep{Jennings2020} tool to remove the azimuthally component of the emission of the disk and leave only the nonaxisymmetric features, as described in the Appendix~\ref{app:alma}. We note that the spiral pattern is much more prominent in the mm continuum than in scattered light. Besides the low S/N of the SPHERE data and the fact that the disk has been reported to be cold, this could also be explained by the anisotropic scattering properties of the dust of different sizes. For a disk that is not edge-on, the larger the particles on the surface layer compared to the wavelength, the more they will scatter light into the disk in forward scattering. Therefore, the amount of light scattered in the line of sight direction would be smaller and would not be detected in scattered light images \citep[e.g.,][]{Mulders2013}. 
    Additionally, we notice that there appears to be a break in the spiral arms on the ALMA image at $\sim$0.16" ($\sim$20\,au), which cannot be observed in our SPHERE data. In the following we treat this break as a separate set of spirals.
    
    In order to characterize the spirals we considered two models. A logarithmic spiral given by:
    \begin{equation}
        r = r_0 \cdot \exp(b\theta),
    \end{equation}
    and an Archimedean spiral, defined as: 
    \begin{equation}
      r = r_0 + b\theta,
    \end{equation}
    
    \noindent where $\theta$ is the polar angle, $r_0$ is the radius for which $\theta$=0, and $b$ relates to the pitch angle ($\mu$) of the spiral. The pitch angle is defined as the angle between the tangents to a spiral arm and a circle drawn from the center of the disk, it describes how tightly the spiral arms are wound. In the logarithmic case, the pitch angle is constant along all radii and it is given by $\mu = \arctan(1/b)$, while for the Archimedean spiral, the pitch angle depends on the radius as $\mu = b/r$. To test the symmetry of the spiral arms, the parameters $r_0$ and $b$ were fitted separately, while we assumed $i$ and PA to be fixed and equal to the literature values shown in Table \ref{tab:wo6}. Therefore, we had four and eight free parameters for the SPHERE and ALMA data, respectively. To fit the data, we used the MCMC code based on \texttt{emcee} \citep{Foreman-Mackey2013} and described in \cite{Kurtovic2018}. A flat prior probability is used for the free parameters. For each fit, we used 250 walkers with two consecutive burning stages of 1000 and 500 steps, and then 1500 steps to sample the parameter space. 
    
    The resulting pitch angles for each fit are given in Table \ref{tab:pitch}, where for the SPHERE data, N corresponds to the northern spiral and S to the southern spiral; and in the case of the ALMA data, N1 corresponds to the northern inner spiral, S1 to the southern inner spiral, N2 to the northern outer spiral and S2 to the southern outer spiral. For the Archimedean fit, the pitch angles are estimated at 35\,au. We find that there are some significant differences between the values of the inner and outer ALMA spirals for the logarithmic model, which leads us to conclude that the two sets might not be part of the same spiral arm. There seems to be additional structure in the region of the discontinuity, however, follow-up deeper observations would be needed to draw any conclusions of the origin of this break. Furthermore, we note that the values of the pitch angle for the corresponding spirals differ from one data set to the other, and that in the case of the Archimedean model, the scattered light pitch angles are slightly higher that those from submillimeter. This can be expected, since we are tracing different regions of the disk (the flared surface vs. the midplane). 
    Moreover, we find that the pitch angles from the Archimedean model decrease with the distance from the star, in agreement with the results of \cite{Huang2018III}.  
    
    \begin{figure}
    \centering
    \includegraphics[width=0.46\textwidth]{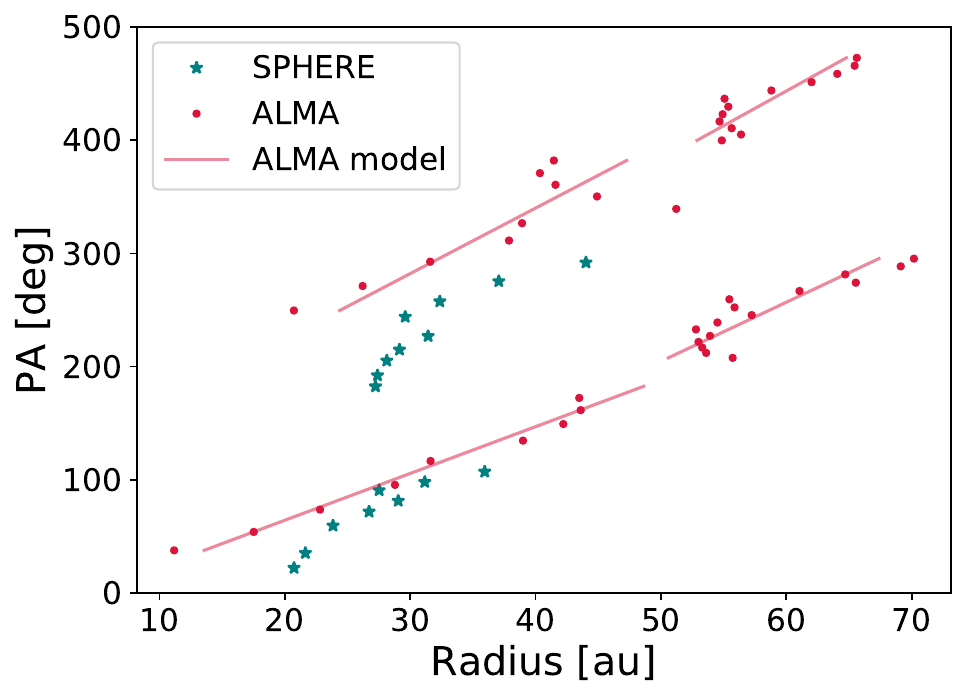}
      \caption{Polar plot showing the spiral arms of both the SPHERE (stars) and ALMA (dots) images of WaOph 6. The lines show the Archimedean best fit model for the spirals.
              }
         \label{fig:polar}
    \end{figure}
    
    Figure \ref{fig:polar} shows the spirals retrieved from both data sets in a polar map, along with the best-fit Archimedean model for each arm. We find a discontinuity in the spiral arms for the ALMA data at $\sim$50 au. This peculiarity has already been reported by \cite{Huang2018III}, who also noted that this discontinuity appears to coincide with a region where there is additional bright emission between the main spiral arms, and comment that it could be explained by either the presence of a ring crossing this region, or "spurs" emerging from the main spirals.
    
    \subsection{Origin of the spirals}
    
    Large perturbations in the disk launch sound waves that result in a spiral shape due to the differential rotation of the disk. Theoretical models have shown that such large perturbations can be driven by the presence of an embedded planetary mass perturber \citep[e.g.,][]{Boccaletti2013}, GI \citep[e.g.,][]{Goldreich1979, Tomida2017}, or a combination of both \citep[e.g.,][]{Pohl2015}.
    
    Two-arm spirals in disks can be driven by a massive, giant planetary companion ($\gtrsim 5 M_J$), that would typically be located at the tip of the primary arm. This scenario suggests, however, that these planets are fainter than predicted by "hot-start" evolutionary models \citep{Dong2018sp}, since the number of detections is low. 
   
   On the other hand, one criteria to test the GI hypothesis is to use the so called \cite{Toomre1964} parameter $Q$, given by
   
   \begin{equation}
    \label{eq:toom}
    \centering
      Q \equiv \frac{\Omega_k c_\mathrm{s}}{\pi G\Sigma},
    \end{equation}
    
    where $\Omega_k$ is the angular velocity, c$_\mathrm{s}$ is the sound speed, and $\Sigma$ is the surface density. These parameters vary with radius in the disk, resulting in $Q$ being a function of the radius. The Toomre stability criteria states that a disk will be gravitationally stable if $Q \geq 1$ and unstable if $Q < 1$.
   
   In the case of WaOph 6, we estimated the Toomre parameter $Q$ using equation \ref{eq:toom} and assuming $\Sigma \propto 1/\sqrt{r}$, where $r$ is the distance from the star, throughout the disk (see Appendix \ref{app:toomre} for the detailed calculation) in order to see whether the disk would be stable under these conditions. We obtained that $Q$ varies from $2.2 -33.2$ from the outer part of the disk inward. This indicates that the disk is fairly stable according to the Toomre stability criterion, implying that a large perturbation driving the spiral arms should have come from a source other than GI. Nonetheless, we are aware that this is not an absolute proof of GI not taking place in the disk at some earlier evolutionary state. This analysis only shows that a disk with a surface density $\propto 1/\sqrt{r}$ around a 0.98 $M_\odot$ star appears to be stable within our assumptions of the disk mass and the gas surface density, which should be taken with caution due to the considerable uncertainties in their calculation.
   In this context, we decide to test the planetary perturber hypothesis by performing hydrodynamical simulations and radiative transfer to compare with our observations.


\begin{figure*}
    \centering
    \begin{subfigure}{0.48\textwidth}
        \centering
        \includegraphics[width=0.95\textwidth]{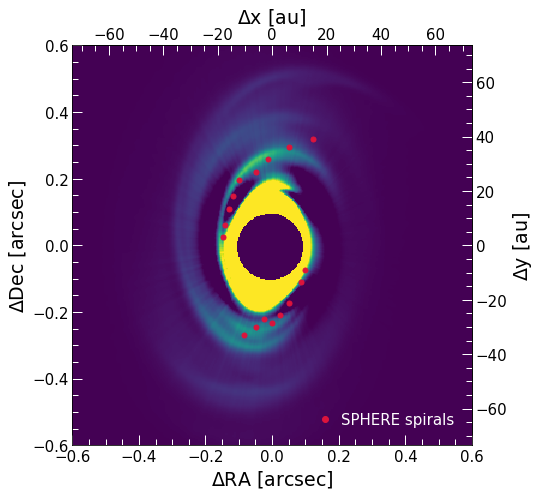}
        \label{fig:sub1}
    \end{subfigure}
    \begin{subfigure}{0.48\textwidth}
        \centering
        \includegraphics[width=0.95\textwidth]{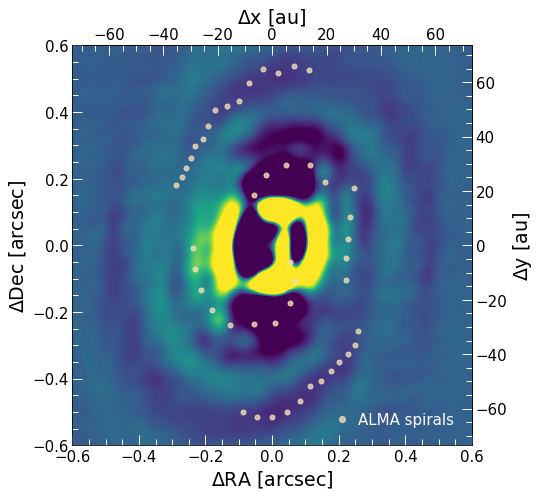}
        \label{fig:sub2}
    \end{subfigure}
\caption{Radiative transfer images showing the spirals formed by a 10\,M$_{\rm{Jup}}$ planet at 140\,au. \textit{Left:} Synthetic polarized scattered light $Q_{\phi}$ image with an analogous Gaussian kernel to the one applied to Fig. \ref{fig:sphspirals}, left. The dark central area shows the coronagraph coverage. \textit{Right:} Synthetic mm continuum image after subtracting the azimuthal average flux on the image plane to enhance the spirals, analogous to the procedure applied to Fig.~\ref{fig:sphspirals}, right. The red and white dots denote the location of the observed spirals for each image, respectively.}
\label{fig:radtr}
\end{figure*}

\section{Modeling}
\label{sec:model}

    In order to test the hypothesis of a forming planet causing the spiral features, we performed 3D gas-only and 2D gas+dust hydrodynamical simulations in which a massive planet at a separations $\geq 90$ au generates large-scale spiral arms interior to its orbit.
    To compare the results to our observations, we fed the resulting density distributions into a radiative transfer code to generate synthetic images.
    
    \subsection{Hydrodynamical models}
   
    To model the dust on the surface sampled by scattered light, we ran 3D simulations of
    the gas dynamics using the hydrodynamical
    code \texttt{PLUTO} \citep{Mignone2007}.
    The gas distribution was initially set
    following a vertically isothermal
    configuration at hydrostatic equilibrium
    \citep[see, e.g.,][]{Fromang2011}.
    The disk temperature depends
    on the cylindrical radius $R$ from the
    center of the domain as $T\propto R^{-1/2}$,
    whereas the gas pressure scale height is
    computed as $H=c_s/\Omega_K$, where
    $c_s\propto T^{1/2}$
    is the local sound speed and
    $\Omega_K\propto R^{-3/2}$ is the
    Keplerian angular velocity. With the
    chosen parameters, the disk aspect ratio
    depends on $R$ as $H/R\propto R^{1/4}$,
    with $H/R=0.1$ at $R=50$ au. The gas
    volumetric density decreases with $R$
    as $\rho\propto R^{-7/4}$, in such
    a way that the surface
    density varies as $\Sigma\approx \sqrt{2 \pi} H \rho \propto R^{-1/2} $. A locally
    isothermal equation of state
    is applied such that the initial temperature 
    distribution is maintained through time.
    
    The hydrodynamical equations were solved
    in spherical coordinates in a reference
    frame centered at the star-planet center of mass corotating with
    the system.
    The computational domain, given by
    the region $(r,\theta,\phi)\in [8,210]
    \,\mathrm{au}\times[\pi/2-0.3,\pi/2+0.3]
    \times[0,2\pi]$, is discretized using
    a grid of resolution $N_r\times N_\theta \times N_\phi=256\times 64\times 512$
    logarithmically spaced in the $r$-direction
    and uniformly spaced in the remaining ones.
    All computed fields are fixed to their initial values at the radial boundaries except for the density
    and the radial velocity, which are
    reflected and extrapolated in the
    ghost zones, respectively. On the
    vertical boundaries, reflective
    conditions are applied.
    The gravitational potential is computed as the sum of the potentials of the star and the planet. To avoid
    divergences, the later is modified
    in the vicinity of the planet location following the 
    prescription by \cite{KlahrKley2006},
    employing a smoothing length equal to
    half the planet's Hill radius. For
    stability purposes, the planet
    mass is smoothly increased from
    $0$ to its final value in a total time of $100$ yr.
    We also included viscosity with constant $\alpha=10^{-3}$. The resulting dust mass distribution was computed assuming a perfect coupling between the dust particles
    and the gas flow, with a uniform dust-to-gas mass ratio of $10^{-2}$.
    
    To model the dust evolution in the midplane sampled mainly by the millimeter observations, we ran two-dimensional hydrodynamical simulations using the multi-fluid version of the code \texttt{FARGO3D}\footnote{\href{http://fargo.in2p3.fr}{http://fargo.in2p3.fr}} \citep{Benitez2016, Benitez2019}. It solves the Navier-Stokes equations of the gas and multiple dust species, each one modeled as a pressureless fluid that represents a specific grain size. We traced eight different dust species in our simulations. 
    The initial gas temperature, gas surface density structure, equation of state and gas viscosity prescription of the 2D model are equivalent to our 3D simulations model, described above. The initial dust surface density in our simulation has the same structure as the gas surface density, while set by an initial dust-to-gas mass ratio of $10^{-2}$ everywhere in the disk. 
    We traced the dynamical evolution of eight dust fluids, that are logarithmically spaced in size, and followed a dust size distribution $n(s)\propto s^{-2.5}$ with minimum and maximum dust sizes of $10\,\rm\mu m$ and $100\,\rm\mu m$. We set the dust intrinsic density to $2.0\,\rm g\,cm^{-3}$. 
    The dynamics of the dust fluids is dictated by its local Stokes number (dimensionless stopping time), defined as $\mathrm{St} = \pi a_i \rho_{int}/2\Sigma_g$, where $a_i$ is the grain size of the size-bin, $\rho_{int}$ the intrinsic grain density and $\Sigma_g$ the gas surface density. Dust diffusion was included in the simulation following the same prescriptions implemented in \cite{Weber2019}, which are based on the results of \cite{Youdin2007}.
    Dust feedback onto the gas, dust growth, and dust fragmentation were not included in our simulations. The two-dimensional grid is linear in azimuth and logarithmic in radius, using 512 cells in $\phi$ covering $2\pi$, and 256 cells in $r$ covering from $\sim8.4$ au to $\sim 420$ au. 
    A planet was slowly introduced over $~ 8 \times 10^{4}$ yr fixed at the given radius, driving spiral density waves in the disk. The planet's potential was smoothed by a length factor of 60\% the disk scale height.
    For a more detailed description of the FARGO3D multi-fluid simulations see also \cite{Weber2019}.
    Our test runs show that only dust fairly well coupled to the gas follows the spiral density waves \citep[as shown by e.g.,][]{Veronesi2019, Sturm2020}, with Stokes numbers below $\sim 10^{-2}$. Dust particles with larger Stokes number decouple from the gas and form axisymmetric rings. Fixing the disk gas surface density given the mass constraint from the observations, and the dust intrinsic density to a standard value, limiting the maximum dust size to $100\,\rm\mu m$ in our models is required to maintain the Stokes number of the dust observed at millimeter wavelengths below $\sim 10^{-2}$, therefore, the simulated dust particles trace the spiral arm structure. A summary of the parameters used in our simulations can be found in Table \ref{tab:simparams}. 
    
    \begin{table}
    \caption[Simulation parameters]{Summary of simulations parameters. } 
    
    	\centering

    	\begin{tabular}{ccc}
    	\hline
        \hline
        	Parameter & 2D gas+dust & 3D gas \\
            \hline
            Aspect Ratio at 100 au  & 0.12 & 0.12 \\
            Flaring Index  & 0.25 & 0.25 \\
            Surface Density Slope & 0.5 & 0.5 \\
            Alpha Viscosity  & $10^{-3}$ & $10^{-3}$ \\
            Stellar Mass & 0.98 $M_{\odot}$  & 0.98 $M_{\odot}$  \\
            Planet-to-Star Mass Ratio  & $10^{-2}$  & $10^{-2}$ \\
            Planet Orbital Radius & 140 au  & 140 au  \\
            $\#$ of Cells in $r$ & 256  & 256  \\
            $\#$ of Cells in $\phi$ & 512 & 512  \\
            $\#$ of Cells in ${\theta}$ & -  & 64   \\
            Grid Inner Radius & 8.4 au  & 8 au  \\
            Grid Outer Radius & 420 au  & 210 au  \\
            Total Evolution Time & $4\times 10^5$ \, $\mathrm{yr}$  & $4\times 10^5$ \,$\mathrm{yr}$  \\
            Dust-to-Gas Mass Ratio  & 10$^{-2}$  & -  \\
            Maximum Dust Size & 100 $\mu$m  & -  \\
            Minimum Dust Size & 10 $\mu$m  & -  \\
            Dust Size Slope & 2.5  & -  \\
            Dust Intrinsic Density & 2.0 g cm$^{-3}$  & -  \\
            \hline
    	\end{tabular} 
    	
    \label{tab:simparams}
    \end{table}
    
    We sample the parameter space of a planet with masses between $2-15$ M$_\mathrm{Jup}$ and separations between $90-160$~au (see Appendix~\ref{app:hydro}, Fig.~\ref{fig:densgall} for some of the resulting density maps). The lower limit in the mass range is chosen based on the results of \cite{Juhasz2015}, who concluded that observable spiral arms are formed for planets with M $ > 1 \mathrm{M}_\mathrm{Jup}$. Tighter constraints on the lower limit of the planet mass can be obtained from spiral arm formation theory. For a planet with a mass larger than three thermal masses (M$_{th}\equiv c_s^3/\Omega G=\mathrm{M}_{\star}(h/r)_p^3$), two spiral arms will form interior to its orbit \citep{Bae2018}. Since we wanted to model $m = 2$ spirals, we chose to place the planets outside of the spiral arm (which extends up to 90 au in the millimeter continuum) based on the results of \cite{Dong2015}. Assuming that the planet is outside $\sim$ 90 au and the disk aspect ratio of our model, we obtained that two spirals are formed for planet masses larger than $\sim$ 4.8 $M_\mathrm{Jup}$.
    Another criteria for the planet mass comes from the separation between the primary and secondary spiral arms ($\phi_{sep}$). \cite{Fung2015} obtained that this quantity scales with the planet mass, following $\phi_{sep}=102^{\circ}(q/0.001)^{0.2}$, where in this case $q$ is the planet-to-star mass ratio. If we consider that the spiral arms have a separation range between $135^{\circ}$ and $180^{\circ}$, we obtain that the planet mass should be between 4 and 17 $M_\mathrm{Jup}$. 
    After a few test runs in our simulations, we realized that in order to observe the disk truncate at 90 au (consistent with the disk outer edge in the millimeter continuum), when increasing the separation, we should also increase the planet mass.
    Snapshots of the gas and total dust surface densities of our 3D and 2D hydro simulations for planets of 5, 10 and 15 M$_\mathrm{Jup}$ at separations of 130, 140 and 160 au, respectively, are shown in the Appendix \ref{app:hydro}.

    \subsection{Radiative transfer}
    
    In order to compare the results generated by the procedure described in the last section with our observations, we generated images in both polarized NIR and millimeter continuum using the radiative transfer code RADMC-3D \citep{Dullemond2012}. 
    We obtained synthetic scattered light images
    using the dust mass distribution computed in the described 3D \texttt{PLUTO} simulations. 
    Based on \cite{Ricci2010},
    we assumed a dust size distribution $n(s)\propto s^{-2.5}$ and model scattering by submicron particles with sizes ranging
    between $0.01$ and $0.5$ $\mu$m. To compute the dust mass in this
    range, we used the total dust mass estimated in this
    work (see Table
    \ref{tab:wo6})
    assuming maximum grain
    sizes in the mm, to obtain
    M$_\mathrm{dust,<0.5\,\mu\mathrm{m}}=10^{-9}$ M$_\odot$.
    Opacities are computed
    assuming a dust composition of
    $60\%$ astronomical silicates
    and $40\%$ amorphous carbon
    grains, taking the optical
    constants respectively from
    \cite{DraineLee1984} and
    \cite{LiGreenberg1997},
    and combining them following
    the Bruggeman mixing rule.
    Scattering matrices were
    computed assuming spherical
    dust grains using the BHMIE
    code \citep{BohrenHuffman1983}
    for Mie scattering.
    For the scattered
    light computations, we
    approximated the grain size
    distribution using $5$
    size bins. To smooth out
    oscillations in the polarization
    degree occurring when considering
    spheres of a single size
    \citep[see, e.g.,][]{Keppler2018},
    we used a Gaussian size
    distribution within each bin
    with a FWHM of $20\%$ of
    the corresponding grain size.
    The star was modeled as a
    point source located at the
    domain center emitting
    thermal radiation with
    characteristics summarized
    in Table~\ref{tab:wo6}.
    We used RADMC-3D to model
    anisotropic scattering
    with full treatment of 
    polarization, using a total
    of $10^8$ photon packages.
    The obtained Stokes $Q$ and $U$
    frames were then convolved by a Gaussian PSF with a FWHM of $51$ mas to reproduce the resolution of the VLT/SPHERE observations (see Section \ref{sec:obs}), after which we used equation \eqref{eq:QphiUphi} to obtain the resulting $Q_\phi$ images.
    
    To compare with the ALMA data, we computed radiative transfer predictions of the dust continuum, in this case, using the output of the dust and gas 2D simulation. We used the dust density field from the simulation as input for RADMC-3D. 
    We expanded the two-dimensional surface density vertically, assuming a Gaussian shape, where the volumetric mass density for each dust bin follows:
    \begin{equation}
        \rho_{i}(r)= \frac{\Sigma_i(r)}{\sqrt{2\pi}H_i(r)} \times \exp\left(-\frac{z^2}{2H_i^2}\right),
    \end{equation}
    where $H_i$ indicates the pressure scale height of the dust bin. The vertical settling of the disk follows a standard diffusion model \citep{Dubrulle1995}:
    \begin{equation}
        H_i=\sqrt {\frac{\tilde{\alpha}}{\tilde{\alpha} + St_i}}H_g,
    \end{equation}
    where $H_g$ is the gas pressure scale height, $St$ is the dust Stokes number, and $\tilde{\alpha}=\alpha/Sc_z$ with $\alpha$ the $\alpha$-viscosity value of the gas. $Sc_z$ is the Schmidt-number, set to 1 $Sc_z$ relates the dust diffusion coefficient with the gas viscosity $D_z=\nu/Sc_z$ (see also \cite{Weber2019}).
    We used optool\footnote{\url{https://github.com/cdominik/optool}} to compute the dust asorption and scattering opacities of a mixture using standard Mie theory and Bruggeman rules.
    We assumed that the composition of the dust grains is a mixture of silicates (internal density of 3.2 g/cm$^3$), amorphous carbon (internal density of 2.3 g/cm$^3$), and vacuum. Assuming that the solids in the mixture are 60\% silicates and 40\% carbon, a volume fraction of 25\% of vacuum in the mixture is required so its internal density is $\sim 2$ g/cm$^3$.
    The dust size distribution is equal to the values used for the simulation, set by the power law $n(s) \propto s^{-2.5}$, with maximum and minimum dust sizes of 100 $\mu$m and 10 $\mu$m, respectively.
    The total dust mass in our models is $\sim 10^{-4}\,\mathrm{M}_{\odot}$. We computed the dust temperature using the Monte Carlo method of \cite{Bjorkman2001}, and the continuum emission image via ray-tracing, taking into account absorption and scattering, assuming Henyey–Greenstein anisotropic scattering. 
    We computed simulated ALMA observations from the radiative transfer synthetic continuum image using CASA (version 5.6) \texttt{simobserve} and \texttt{tclean} tasks. Following the observations setup from the DSHARP survey \citep{Andrews2018}, we simulated an 8 h integration in configuration C43-8 combined with a 15 min integration in C43-5. Finally, we cleaned the image using briggs weighting 1.0. We obtained a beam size of $55\times53$ mas and PA of $\sim -55^{\circ}$, directly comparable to the ALMA observation.
    
    \subsection{Results and comparison to observations}
    \label{sec:simres}
    
     All tested planets drive 
    $m=2$ spiral arms whose
    symmetry increases for
    larger planetary mass and have a low contrast in the dust surface density (as seen in the density plots shown in Fig.~\ref{fig:densgall}). Given the
    asymmetry in the $5$ M$_\mathrm{Jup}$ case,
    we conclude that in case
    the spirals are caused by
    a planet, its mass should
    be at least of approximately $10$ M$_\mathrm{Jup}$.
     In Fig.~\ref{fig:radtr} we show resulting radiative transfer images for a 10 M$_\mathrm{Jup}$ planet at a separation of 140 au, with
     spiral arms observable both in the scattered light and millimeter continuum observations.
     For a better comparison to the simulations, we apply a Gaussian kernel to the image on the left panel, similar to the one used for the SPHERE image in Fig.~\ref{fig:sphspirals}, left; and we subtract the azimuthal average flux on the image plane to enhance the spirals on the synthetic millimeter continuum image on the right, analogous to the procedure applied to Fig.~\ref{fig:sphspirals}, right. Additionally, we overplot the location of the observed spiral arms.
     The obtained images
     resemble the ones detected
     both in the scattered light
     and millimeter continuum
     observations, except for the fact that we are only able to fit either the inner or the outer spirals from the millimeter observations, but not both at the same time (see Fig.~\ref{fig:gall}, lower panel). 
    This is likely due to
    missing physics in our
    simulations, as these models
    of spirals launched
    by a single planet
    are unable to
    reproduce the break
    in the spirals observed
    by ALMA, as well as the gap and the ring features (at 79 and 88 au, respectively) in the observations.
     We must also note that in order to see spiral arms induced by a planet in millimeter continuum, the dust must be fragmentation limited \citep[e.g.,][]{Birnstiel2010} leading to a small dust maximum size, and therefore, to Stokes numbers small enough to follow the spirals. In protoplanetary disks, the maximum grain size is mainly set by radial drift or fragmentation of particles after collisions. The later depends on the disk viscosity and the threshold considered for the fragmentation velocity of the grains. Assuming low fragmentation velocities for ice grains \citep[e.g., $<$ 1 m/s, as suggested by recent laboratory experiments such as][]{Musiolik2019, Steinpilz2019}, and alpha=$10^-3$ (as taken in the simulations), the maximum grain size in the entire disk is dominated by fragmentation, limiting the maximum size of ~100$\mu$m \citep{Pinilla2021}. \cite{Kataoka2016} have found that dust with similar characteristics is traced by millimeter continuum observations of the similarly young disk HL Tau. These characteristics are not required to see spirals generated by GI, where the dust trapping in spirals is efficient for larger dust Stokes numbers \citep{Rice2004}. 
    We also note that none of the parameter sets that we sample are able to reproduce the contrast nor the apparent break in the spiral arms shown in the ALMA data, which might be explained by additional physical processes occurring in the disk. However, more complex simulations including other effects (e.g., dust growth, fragmentation, dust feedback, gas temperature evolution) are beyond the scope of this work. Additionally, we note that a planet of 10 M$_\mathrm{Jup}$ in such a young disk could have either formed via gravitational collapse when the disk was probably more massive and, therefore, gravitationally unstable \citep{Boss1997}, or formed as a stellar companion from cloud fragmentation due to the planet/star mass ratio \citep[$\sim$1\%,][]{Raggiani2016}. 
    We would like to mention that this is a first attempt to find a plausible planetary model to explain the observed spiral pattern in the protoplanetary disk around WaOph~6 and that further, deeper observations would be needed to confirm or discard this scenario.
    
    Since we employ an isothermal equation of state, the spirals produced in our simulations are induced solely by Lindblad resonances and not by buoyancy modes, which may be triggered when using finite cooling times. It is argued in \cite{Bae2021} that such modes cannot be observed in millimeter continuum observations, but could potentially be seen in scattered light. The pitch angles for buoyancy resonances shown in that work for up to 2 M$_\mathrm{Jup}$ planets are generally below those seen in our SPHERE observations (see Table \ref{tab:pitch}), which suggests that the observed spirals are likely not triggered by such modes. Future resolved CO line emission observations analyzing the disk kinematic structure could help discard or verify this hypothesis \citep{Bae2021}.


\section{Discussion}
\label{sec:disc}

\subsection{Observations in NIR and mm}

 Our SPHERE/IRDIS-DPI observations show the launch of an $m=2$ spiral pattern in the disk around WaOph~6. This is a surprising finding, since so far, no spiral arms had been observed in scattered light in disks around K and/or M stars with ages $<1$\,Myr. Moreover, spiral arms have not been observed at these wavelengths in single T Tauri stars of any age \citep{Garufi2018}. Disks with spiral arms detected in scattered light are thought to be older (with the caveat that stellar ages are highly uncertain), and with stellar hosts of spectral types from G to A (e.g., MWC 758, \cite{Dong2018}, HD 142527, \cite{Claudi2019}, HD 100546, \cite{Perez2020}, AB Aur, \cite{Boccaletti2020}, HD 100453, \cite{Benisty2017}). In the millimeter continuum, most of these disks show asymmetric morphologies, along with large cavities \citep[e.g.,][]{Tang2017, Cazzoletti2018, Pineda2019}. Further observations in both scattered light and millimeter continuum of K and M type stars with disks would be needed to determine whether spiral arms are a common feature in such young disks, as well as the possible implications that this might have in dust and gas evolutionary models. We also note that comparing observations at different wavelengths can contribute greatly to the understanding of the physical processes driving the different morphologies seen in protoplanetary disks. 

From our hydrodynamical simulations, we observe that in order to obtain a spiral pattern that can be observed in the millimeter continuum data, the dust particles must have a limited maximum size. This has previously been observed in dust evolution simulations by \cite{Gerbig2019}, and can be linked to the young age of the disk. 

\begin{figure}
    \centering
    \includegraphics[width=0.5\textwidth]{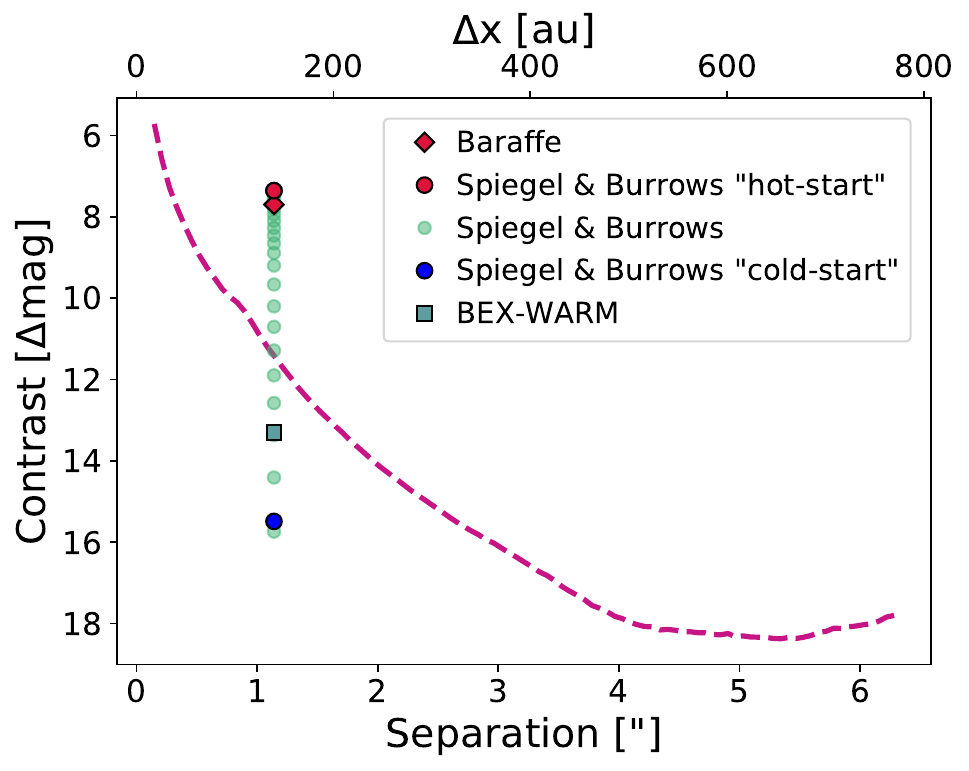}
      \caption{Planet detection limits as a function of the separation from the star for the SPHERE H-band. The purple curve is the $3\sigma$ contrast obtained from the total intensity SPHERE image of WaOph~6. The markers show the magnitude contrast of the proposed 10 M$_\mathrm{Jup}$ planet at 140 au estimated from different evolutionary models. The red markers show the resulting contrast for the ``hot-start'' scenario from the \protect\cite{Baraffe2003} (red diamond) and \protect\cite{Spiegel2012} (red dot) models, while the blue dot shows the contrast for a ``cold-start'' from the \protect\cite{Spiegel2012} models. The green square shows the contrast from the BEX-WARM models (see text). And the green dots show the contrast for different initial entropy values from the \protect\cite{Spiegel2012} models.
              }
    \label{fig:contrast}
   \end{figure}

\subsection{Upper limits on the brightness of point sources}

We used the total intensity image derived from our SPHERE/IRDIS-DPI observations to obtain information on the detection limits for WaOph~6. We built the contrast curve in Fig. \ref{fig:contrast} by considering the contrast between WaOph~6 (the central brightest pixel) and a representative planetary signal in the total intensity image. We took the planetary signal to be three times the noise (root mean square) in 2 pixel wide annuli centered on the star, at different separations up to $\sim$6" ($\sim$740 au). Additionally, we estimated the foreground extinction in the H-band toward WaOph~6. For this we first estimated the reddening by using the intrinsic $J-H$ magnitude of a K6V star from \cite{Pecaut2013}, then using the values in Table 3 of \cite{Rieke1985}, we obtained a visual extinction of $A_V = 5.08$ mag, and an H-band extinction of $A_H = 0.88$ mag.

To estimate the apparent magnitude of our proposed planet, we used two independent evolutionary model predictions. On one hand we considered the evolutionary models by \cite{Baraffe2003} for a 10 M$_\mathrm{Jup}$ planet at 1\,Myr. 
On the other hand, we used the evolutionary models proposed by \cite{Spiegel2012} for both a ``hot'' and ``cold-start'' scenarios, and we extrapolated the H-band absolute magnitude (M$_H$) for our 10 M$_\mathrm{Jup}$ planet at 0.7\,Myr. 
Considering extinction toward WaOph~6, we obtain $\mathrm{m}_H = 15.26$ mag, $\mathrm{m}_H = 14.91$ mag, and $\mathrm{m}_H = 23.07$ mag, respectively for each model. Finally, with the H-band magnitude for WaOph~6, $\mathrm{m}_H = 7.57$ mag, we obtained the following contrasts: $\Delta$mag$= 7.70$, $\Delta$mag$= 7.36$, and $\Delta$mag$= 15.49$ mag, respectively. Furthermore, we obtained the contrasts for our proposed planet in the ``warm-start'' scenario from the initial entropy values reported by \cite{Spiegel2012}. And as an additional comparison, we used the Bern EXoplanet cooling curves \citep[BEX,][]{Mordasini2017} coupled with the COND atmospheric models \citep{Allard2001} reproducing the cooling under ``warm-start'' initial conditions \citep{Marleau2019}, and thus denominated BEX-WARM model \citep[see][and references therein for more details]{Asensio-Torres2021}. As seen in Fig. \ref{fig:contrast}, the detectability of our proposed planet strongly depends on the adopted formation model. In case of the ``hot-start'' scenario, the planet should have been observed, while for a large part of the ``warm-start'' and for the ``cold-start'' scenarios, the planet contrast is below our detection limits. Based on the planet mass and location, a ``warm'' to ``cold'' start model would be more plausible to explain its existence. 

Additional detection limits for WaOph 6 in the L'-band ($\lambda_0=3.8\,\mu$m) have been recently reported by \cite{Jorquera2020}. They do not detect any companion candidates to the star, but report detection probability maps obtained using the \citep{Baraffe2003} models. From these they preliminary rule out the presence of companions with masses $> 5$~M$_\mathrm{Jup}$ at separations $> 100$~au. However, they advise that these estimates might be optimistic, since they do not consider extinction effects, either toward WaOph~6, nor due to the disk dust. An additional caveat comes from the models, as they become very uncertain in accurately predicting the properties of very young planets. 
It is also important to note that our hydrodynamical simulations do not include additional physical processes that could be ongoing in the disk, coming from the fact that spiral arm formation by a planetary mass object is still not well understood. This could lead to an overestimation of the planet mass, which along with evolutionary models uncertainties, could explain our differing results.


\section{Summary and conclusions}
\label{sec:sum}

We have presented for the first time scattered light SPHERE/IRDIS-DPI observations of the disk around WaOph~6 in the H-band. We analyzed the disk morphology, and used archival ALMA data to compare with ours. We tested the planetary mass perturber hypothesis as the underlying cause for the spiral structure by performing hydrodynamical simulations and using radiative transfer. Our results are summarized below:

\begin{enumerate}
    \item We observe the launch of a set of $m=2$ spiral arms up to $\sim$0.3" (40 au) in our $Q_\phi$ SPHERE/IRDIS-DPI images as seen in Fig. \ref{fig:finalqphi}, left. These spirals were first detected using millimeter continuum observations from the ALMA/DSHARP survey. 
    To our knowledge, WaOph 6 is the youngest disk to show spiral features in scattered light \citep{Garufi2018}. We note that this might be of interest for dust and gas evolutionary models. \\
    \item We observe a companion candidate at about 3" from the star in our data, as shown in the top panel of Fig. \ref{fig:cc}. After the astrometric analysis described in Section \ref{sec:cc}, we were able to determine that the CC is not bound to WaOph~6. With this we also discard the CC being a possible cause of the spiral structure. \\
    \item Comparing our SPHERE observations with archival ALMA/DSHARP data, we find that both the gap and the ring features at 79 and 88 au, respectively, seem to be present in both data sets. We traced the spiral features in both observations as seen in Fig. \ref{fig:sphspirals}. 
    For the ALMA data, we notice a break in the spiral arms of the ALMA image at $\sim$0.16" ($\sim$20 au), which is not observed in our SPHERE data. We treated this break as a separate set of spirals, however, its origin remains unknown. When plotting the spirals in polar coordinates (Fig. \ref{fig:polar}) we find a discontinuity in the spiral arms for the ALMA data at $\sim$50 au, already reported by \cite{Huang2018III}. \\
    \item To test the planetary mass perturber hypothesis we performed hydrodynamical simulations combined with radiative transfer to compare with the observations. We tested the parameter space of a planet with masses between
    $2-15$ $\mathrm{M}_{\rm{Jup}}$ and separations between $90-160$ au (i.e., outside of the spiral structure). All tested planets drive $m=2$ spiral arms. However, none of the parameter sets that we sample are able to reproduce the contrast nor the apparent break in the spiral arms shown in the ALMA data, which may be due to additional physical processes occurring in the disk. Furthermore, the tested planets do not reproduce the gap nor the ring features at 79 and 88 au, respectively, these features need further investigation outside the scope of this work. Given the symmetry of the observed spirals, we find that, if these are caused by a planet, its mass is likely of at least 10~M$_{\rm{Jup}}$. This is a first attempt to explain the spiral structure seen in both data sets, and more data are needed to better constrain the underlying cause of the spiral features. \\ 
    \item To determine the sensitivity of our data to possible companions embedded in the disk, we generated the contrast curve in Fig. \ref{fig:contrast} from the total intensity image. With this we obtain contrast limits for a planetary/substellar companion forming inside the disk in polarized light. We estimate the contrast of our proposed planet using different evolutionary models, where the possibility of detection strongly depends on the formation scenario. A ``warm'' to ``cold'' starts would explain the nondetection of the planet in our SPHERE data. 
\end{enumerate}

In conclusion, the findings in this work highlight the still unknown complexity of WaOph~6. The striking presence of a spiral pattern in scattered light even in limited S/N data are worth further, deeper observations of this source. Which will additionally serve to confirm or discard a planetary perturber as a possible cause behind the spiral features.

\begin{acknowledgements}
SPHERE is an instrument designed and built by a consortium consisting of IPAG (Grenoble, France), MPIA (Heidelberg, Germany), LAM (Marseille, France), LESIA (Paris, France), Laboratoire Lagrange (Nice, France), INAF - Osservatorio di Padova (Italy), Observatoire de Gen\`{e}ve (Switzerland), ETH Z\"{u}rich (Switzerland), NOVA (Netherlands), ON ERA (France) and ASTRON (Netherlands) in collaboration with ESO. SPHERE was funded by ESO, with additional contributions from CNRS (France), MPIA (Germany), INAF (Italy), FINES (Switzerland) and NOVA (Netherlands). SPHERE also received funding from the European Commission Sixth and Seventh Framework Programmes as part of the Optical Infrared Coordination Net- work for Astronomy (OPTICON) under grant number RII3-Ct-2004-001566 for FP6 (2004-2008), grant number 226604 for FP7 (2009-2012) and grant number 312430 for FP7 (2013-2016). This work has made use of the SPHERE Data Centre, jointly operated by OSUG/IPAG (Grenoble), PYTHEAS/LAM/CeSAM (Marseille), OCA/Lagrange (Nice), Observatoire de Paris/LESIA (Paris), and Observatoire de Lyon (OSUL/CRAL).This work is supported by the French Na- tional Research Agency in the framework of the Investissements d’Avenir program (ANR-15-IDEX-02), through the funding of the "Origin of Life" project of the Univ. Grenoble-Alpes. This work is jointly supported by the French National Programms (PNP and PNPS). AV acknowledges funding from the European Research Council (ERC) under the European Union’s Horizon 2020 research and innovation programme (grant agreement No. 757561). A-M Lagrange acknowledges funding from French National Research Agency (GIPSE project).\\

Paola Pinilla. and Nicol\'as T. Kurtovic acknowledge support provided by the Alexander von Humboldt Foundation in the framework of the Sofja Kovalevskaja Award endowed by the Federal Ministry of Education and Research.\\

The research of Julio David Melon Fuksman and Hubert Klahr is supported by the German Science Foundation (DFG) under the priority program SPP 1992: “Exoplanet Diversity” under contracts KL 1469/16-1/2. \\

M. Barraza-Alfaro acknowledges funding from the European Research Council (ERC) under the European Union’s Horizon 2020 research and innovation program (grant agreement No. 757957).\\

P.W. acknowledges support from ALMA-ANID postdoctoral fellowship 31180050.\\

This paper makes use of the following ALMA data: \href{https://almascience.nrao.edu/aq/?result_view=observation&projectCode=2016.1.00484.L}{ADS/JAO.ALMA\#2016.1.00484.L}. ALMA is a partnership of ESO (representing its member states), NSF (USA) and NINS (Japan), together with NRC (Canada), MOST and ASIAA (Taiwan), and KASI (Republic of Korea), in cooperation with the Republic of Chile. The Joint ALMA Observatory is operated by ESO, AUI/NRAO and NAOJ.

\end{acknowledgements}

%
%

\bibliographystyle{aa} 
\bibliography{biblio} 

\begin{appendix} 
\section{Dust mass estimate}
\label{app:a}
   
   Millimeter continuum observations, obtained assuming optically thin emission \citep{Hildebrand1983}, allow us to use the relation

\begin{equation}
\label{eq:mass}
    \centering
      M_\mathrm{dust} \simeq \frac{d^2F_{\nu}}{\kappa_{\nu}B_{\nu}(T(r))},
\end{equation}

where $d$ is the distance to the star; $F_{\nu}$ is the total flux at a given frequency $\nu$; $\kappa_{\nu}$ is the dust opacity at a given frequency, for which we used the common relation applied to disk surveys, $\kappa_{\nu}$ = 2.3 $\mathrm{cm^2 g^{-1}} \times$ ($\nu$/230 GHz)$^{0.4}$ \citep{Andrews2013}; and $B_{\nu}(T_\mathrm{dust})$ is the Planck function for a given dust temperature $T_\mathrm{dust}$, that we derived from the relation

\begin{equation}
    \centering
      T_\mathrm{dust} = 22 \times (L_*/L_{\astrosun})^{0.16} K,
\end{equation}

from \cite{vanderPlas2016}, which gives $T_\mathrm{dust} =$ 26.05 K. The resulting dust mass from equation \ref{eq:mass} is reported in Table \ref{tab:wo6}, and, assuming a dust/gas mass ratio ($M_\mathrm{dust}/M_\mathrm{gas}$) of 1:100, within the previously reported values. However, we are aware that the assumptions made to perform this calculation could significantly differ from the actual disk conditions and therefore, this result should be taken with caution. 

\section{Unprocessed reduced $Q_{\phi}$ and $U_{\phi}$ SPHERE images}
\label{app:sph_img}

\begin{figure*}
\includegraphics[width=0.95\textwidth]{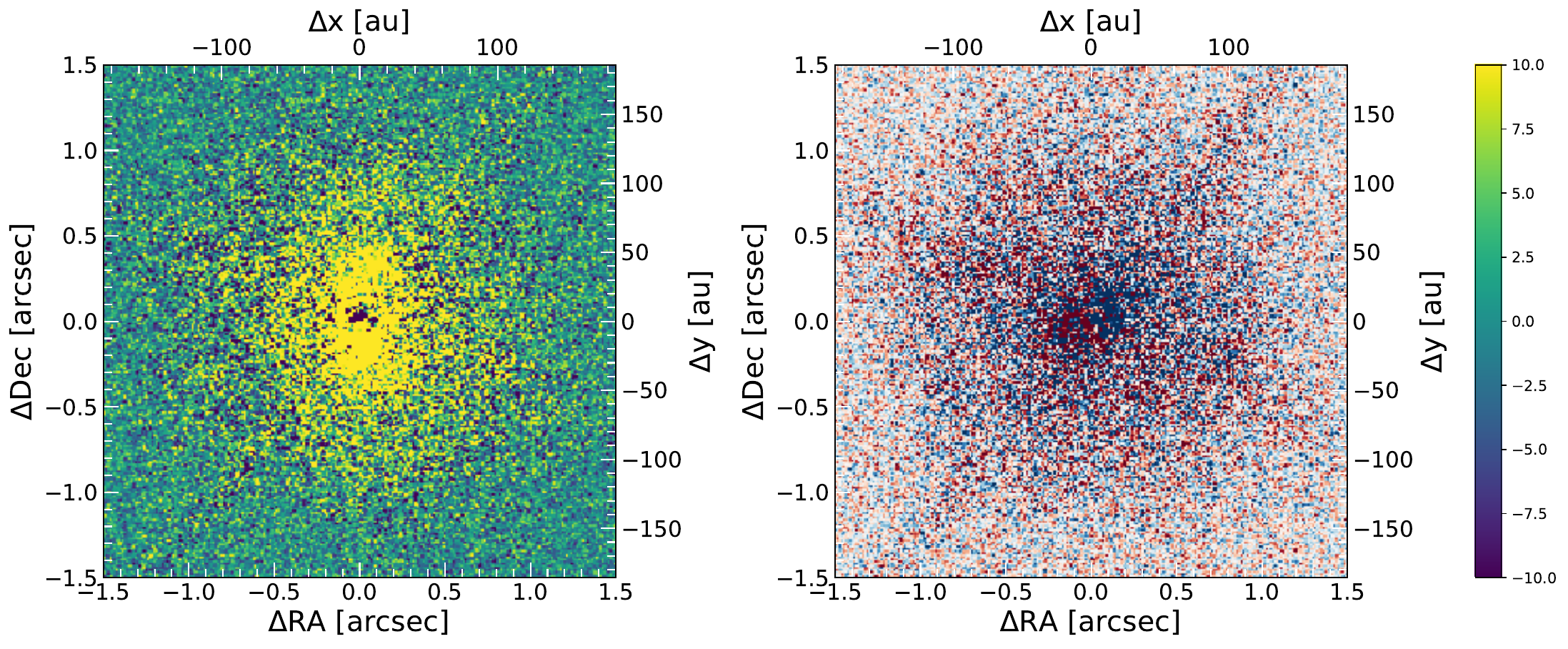}
\caption{Reduced $Q_{\phi}$ and $U_{\phi}$ SPHERE/IRDIS-DPI images. The images are shown up to the distance where the noise dominates. Most of the signal is contained in the $Q_{\phi}$ image.}
\label{sph_red}
\end{figure*}

Figure~\ref{sph_red} shows the reduced $Q_{\phi}$ and $U_{\phi}$ images (on the left and right panels, respectively). The raw data was reduced as detailed in Section~\ref{sec:obs}. Most of the signal is concentrated in the $Q_{\phi}$ image. Due to the low S/N, these images had to be processed for the analysis as described in Section~\ref{sec:sph_obs}.

\section{On extracting the nonaxisymmetric information from the ALMA data}
\label{app:alma}

To recover the millimeter spiral structure, we follow a similar
procedure to the one described in the Appendix B of \cite{Isella2019}. We start from the calibrated visibilities of the dust continuum emission, available from the DSHARP data release. We run a MCMC (Monte Carlo Markov Chain) with 50 walkers to find the offset ($\delta$\,RA,
$\delta$\,Dec) that minimizes the imaginary part of the visibilities,
this gives us the centroid of the disk. In this MCMC we use a flat prior over both dimensions. After correcting by that center, we use the
inclination and position angle measured by \cite{Huang2018} to deproject the visibilities. Our new deprojected data set is analyzed with \texttt{frank} \citep{Jennings2020}, and the best visibilities profile found by this package is substracted from our deprojected data set. The result is a visibility set which only contains the nonaxisymmetric information of the disk, shown in the right panel of Figure \ref{fig:sphspirals}.

\section{Toomre parameter calculation} 
\label{app:toomre}

From equation \ref{eq:toom}, we take 

    \begin{equation*}
    \label{eq:om}
    \centering
      \Omega_k = (GM_*/r^3)^{1/2},
    \end{equation*}
    
    \begin{equation*}
    \label{eq:cs}
    \centering
      c_s = h\Omega_k,
    \end{equation*}
    
    where $h \propto r^{5/4}$, and 
    
    \begin{equation*}
    \label{eq:sig}
    \centering
      \Sigma = \Sigma_0 r^{-1/2},
    \end{equation*}
    
    where 
    
    \begin{equation*}
    \label{eq:sig}
    \centering
      \Sigma_0 = \frac{3M_\mathrm{disk}}{4\pi}\frac{1}{r_\mathrm{max}^{3/2}-r_\mathrm{min}^{3/2}},
    \end{equation*}
    
    which finally leads to $Q \propto r^{-5/4}$. We use $r_\mathrm{min} = 20$ au due to the inner working angle limit of the observations, and $r_\mathrm{max} = 175$ au as the outer radius from the lower limit value used by \cite{Ricci2010}, the only difference when taking the upper limit is that the disk becomes unstable by $\sim$ 330 au.

\section{Gallery of density distributions from the hydrodynamical simulations and radiative transfer images}
\label{app:hydro}

\begin{figure*}
\centering
\includegraphics[width=\textwidth]{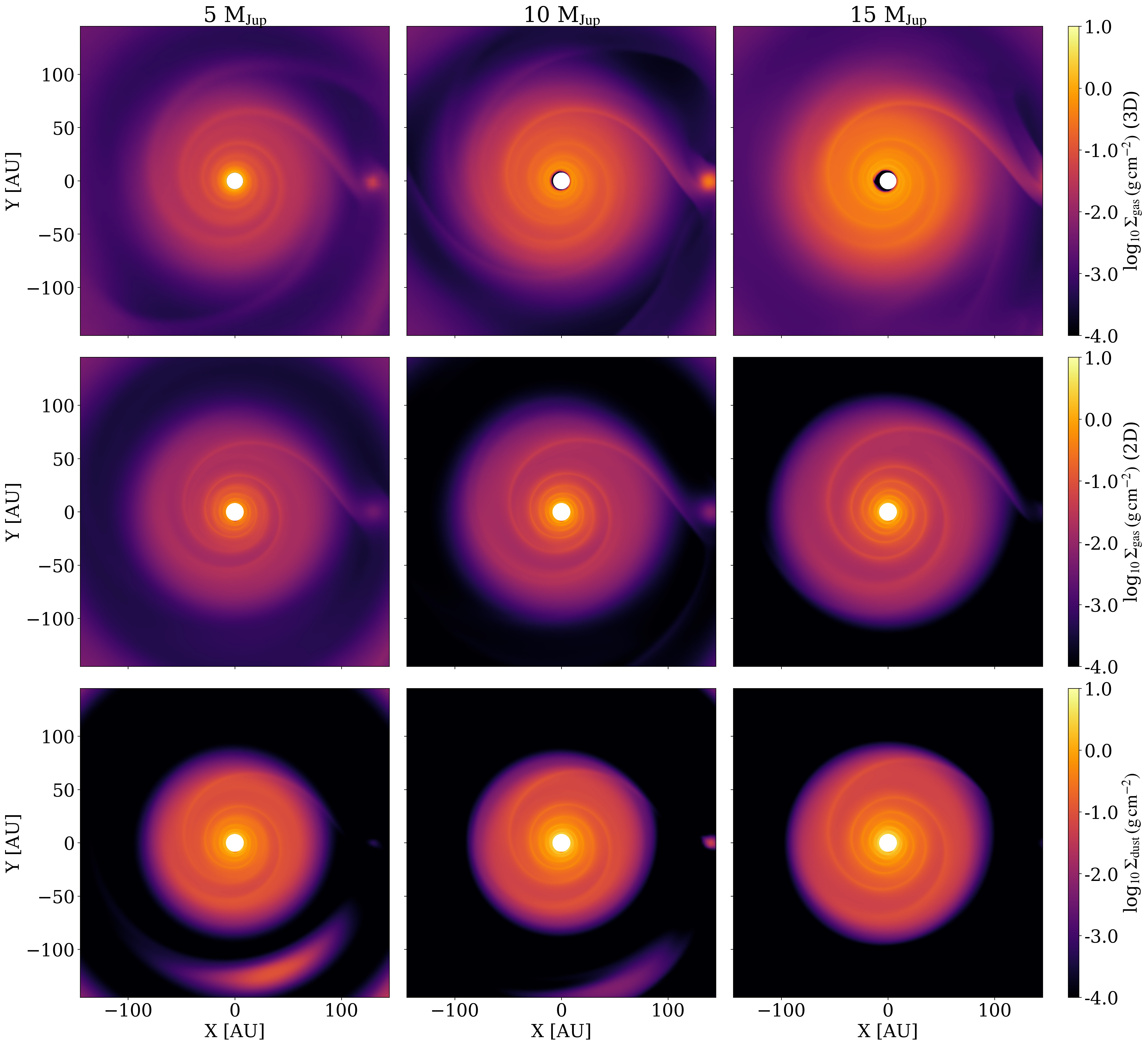}
\caption{Density maps from our 3D (top panels) and 2D (middle and bottom panels) hydrodynamical simulations for planets of 5, 10 and 15 M$_\mathrm{Jup}$ at separations of 130, 140 and 160 au, respectively, shown from left to right. The top and middle panels show the gas surface density maps, while the bottom panels show the dust density maps.}
\label{fig:densgall}
\end{figure*}

\begin{figure*}
\centering
\includegraphics[width=1\textwidth]{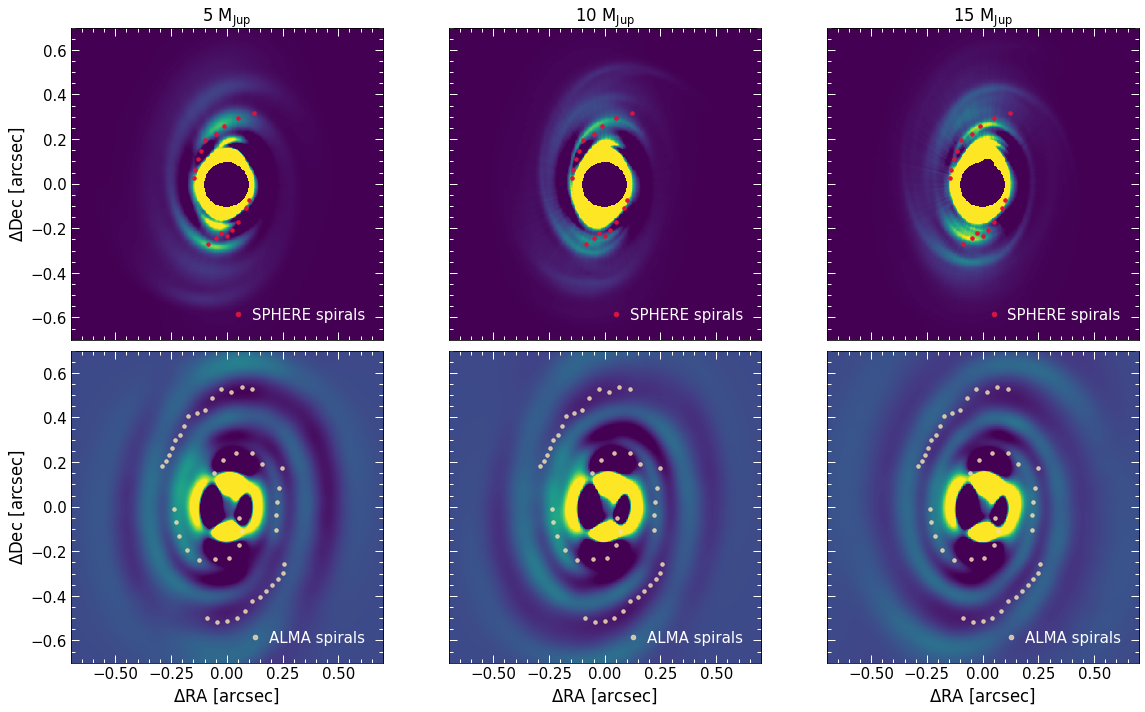}
\caption{Resulting radiative transfer images from our 3D (upper panels) and 2D (lower panels) hydrodynamical simulations for the planets described in Fig.~\ref{fig:densgall}. The mass and separation increase from left to right. The images have been processed with the same techniques as the ones in Fig.~\ref{fig:radtr} for a better comparison to the observations. For the lower panels, we show the images that fit the inner spirals.}
\label{fig:gall}
\end{figure*}

Density maps from our 3D gas and 2D gas $+$ dust hydrodynamical simulations for a planet of 5, 10 and 15 M$_\mathrm{Jup}$ at separations of 130, 140 and 160 au, respectively are shown in Fig. \ref{fig:densgall}. The resulting radiative transfer images from these simulations are shown in Fig.~\ref{fig:gall}. For the case of the synthetic ALMA images, our simulations do not fit the inner and outer spirals at the same time (see Section~\ref{sec:simres}), we show the ones fitting the inner spirals.

\end{appendix}

\end{document}